\documentclass[useAMS]{mn2e}
\usepackage{times,epsfig} 

\title[On the evolution of cooling cores]{On the evolution of cooling cores
in X--ray galaxy clusters}

\author[]
{\parbox[]{6.in} {S. Ettori$^1$, F. Brighenti$^2$ \\
\footnotesize
$^1$ INAF, Osservatorio Astronomico di Bologna, via Ranzani 1,
  I-40127 Bologna, Italy (stefano.ettori@oabo.inaf.it) \\
$^2$ Dipartimento di Astronomia, Universit\`a di Bologna, via Ranzani 1,
  I-40127 Bologna, Italy (fabrizio.brighenti@unibo.it)
}}                                            
\date{}

\begin{document}
\maketitle

\begin{abstract}
To define a framework for the formation and evolution of the cooling cores
in X-ray galaxy clusters, we study how the physical properties change
as function of the cosmic time in the inner regions of a 4 keV and
8 keV galaxy cluster under the action of radiative cooling and gravity only.
The cooling radius, $R_{\rm cool}$, defined as the radius at which the 
cooling time equals the Universe age at given redshift, evolves 
from $\sim 0.01 R_{200}$ at $z>2$, where the structures begin their evolution, 
to $\sim 0.05 R_{200}$ at $z=0$. The values measured at $0.01 R_{200}$ 
show an increase of about 15-20 per cent per Gyr 
in the gas density and surface brightness and a decrease with a mean rate 
of 10 per cent per Gyr in the gas temperature.
The emission-weighted temperature diminishes by about 25 per cent and the 
bolometric X-ray luminosity rises by a factor $\sim2$ after 10 Gyrs when all the
cluster emission is considered in the computation. On the contrary, 
when the core region within $0.15 R_{500}$ is excluded, the gas temperature value
does not change and the X-ray luminosity varies by $10-20$ per cent only.
The cooling time and gas entropy radial profiles are well represented by
power-law functions, $t_{\rm cool}=t_0 +t_{0.01} (r / 0.01 R_{200})^{\gamma}$
and $K=K_0 +K_{0.1} (r / 0.1 R_{200})^{\alpha}$, with $t_0$ and $K_0$ that 
decrease with time from 13.4 Gyrs and 270 keV cm$^2$ in the hot
system (8.6 Gyr and 120 keV cm$^2$ in the cool one) and reach zero 
after about 8 (3) Gyrs.
The slopes vary slightly with the age, with $\gamma \approx 1.3$ and 
$\alpha \approx 1.1$.
The behaviour of the inner slopes of the gas temperature and density profiles
are the most sensitive and unambiguous tracers of an evolving cooling core.
Their values after 10 Gyrs of radiative losses, $T_{\rm gas} \propto r^{0.4}$ and
$n_{\rm gas} \propto r^{-1.2}$,
are remarkably in agreement with the observational constraints available for nearby
X-ray luminous cooling core clusters. Because our simulations do not consider
any AGN heating, they imply that the feedback process does not greatly alter
the gas density and temperature profiles as generated by radiative cooling 
alone. 
\end{abstract}


\section{INTRODUCTION} 

The hot plasma in groups and clusters of galaxies 
during the hierarchical formation of dark matter halos cools 
by bremsstrahlung and line emission regulated by $n_{\rm gas}^2$.
The action of the cooling is thus more effective in the central regions
at higher density, with a twofold effect of 
decreasing the temperature and rising the density moving inwards.
In recent years, the new generation of X-ray telescopes like
{\it Chandra} and {\it XMM-Newton} has been able to spatially resolve
the emission from the central regions characterizing 
the physical properties of the X-ray emitting plasma.
These observations have revealed that the cooling is not
the only mechanism that is responsible for the energetic
of the central gas. Still-debated feedback processes have
to play a relevant role in its thermal evolution
(e.g. Peterson \& Fabian 2006). 

A further unknown property is how a cool core evolve with cosmic time.
Bauer et al. (2005) find that cool cores appear to be common at redshift
$0.15-0.4$, even though there is mounting evidence that they are 
less numerous and/or prominent at higher redshift (Ettori et al. 2004, 
Vikhlinin et al. 2006b, Santos et al. 2008).
In the present work, we try to address this issue from a theoretical point
of view, just considering how a plasma
in hydrostatic equilibrium with a NFW potential with global quantities
consistent with present-day observational constraints,
changes its physical properties as function of the cosmic time under 
the action of the radiative cooling only.
This simple model allows to define a framework in which we can 
make zero-th order predictions on the formation, evolution and fate 
of a cooling core in X-ray groups and clusters of galaxies, taking
into consideration that any deviation from these expectations
has to refer to the variety of the heating mechanisms 
that can affect the energy budget of the structure.

\begin{figure*}
\hbox{
 \epsfig{figure=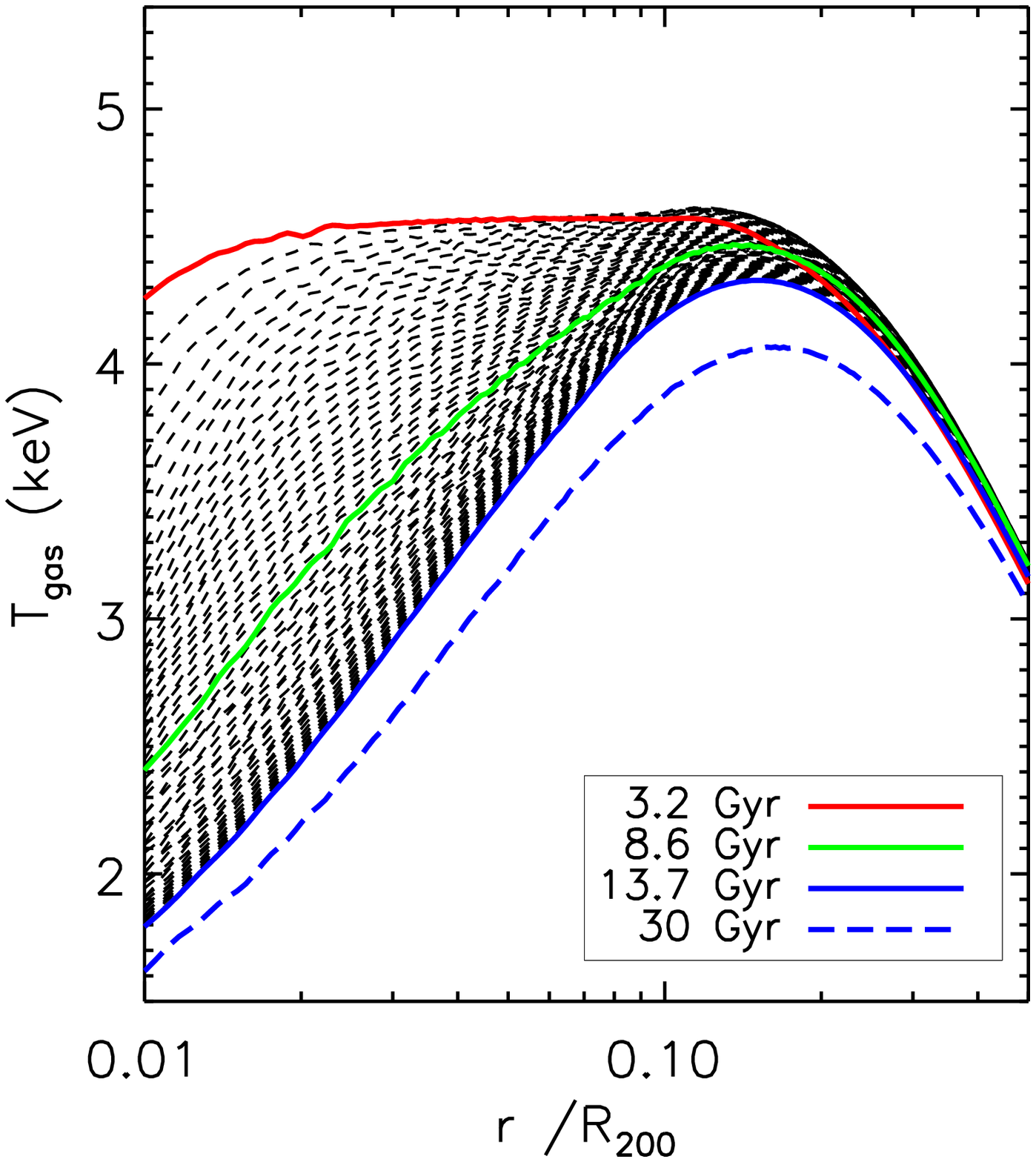,width=0.25\textwidth}
 \epsfig{figure=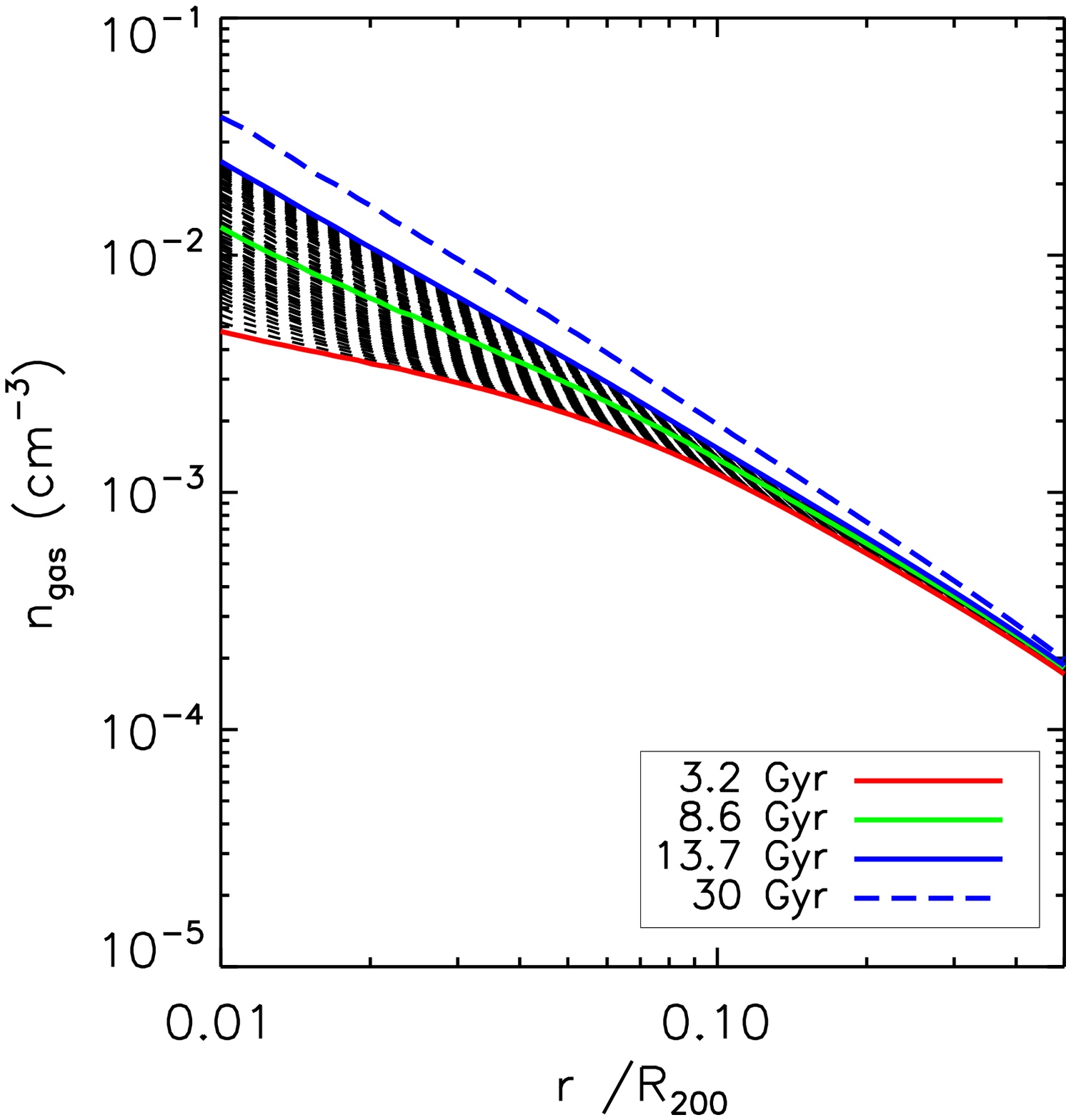,width=0.25\textwidth}
 \epsfig{figure=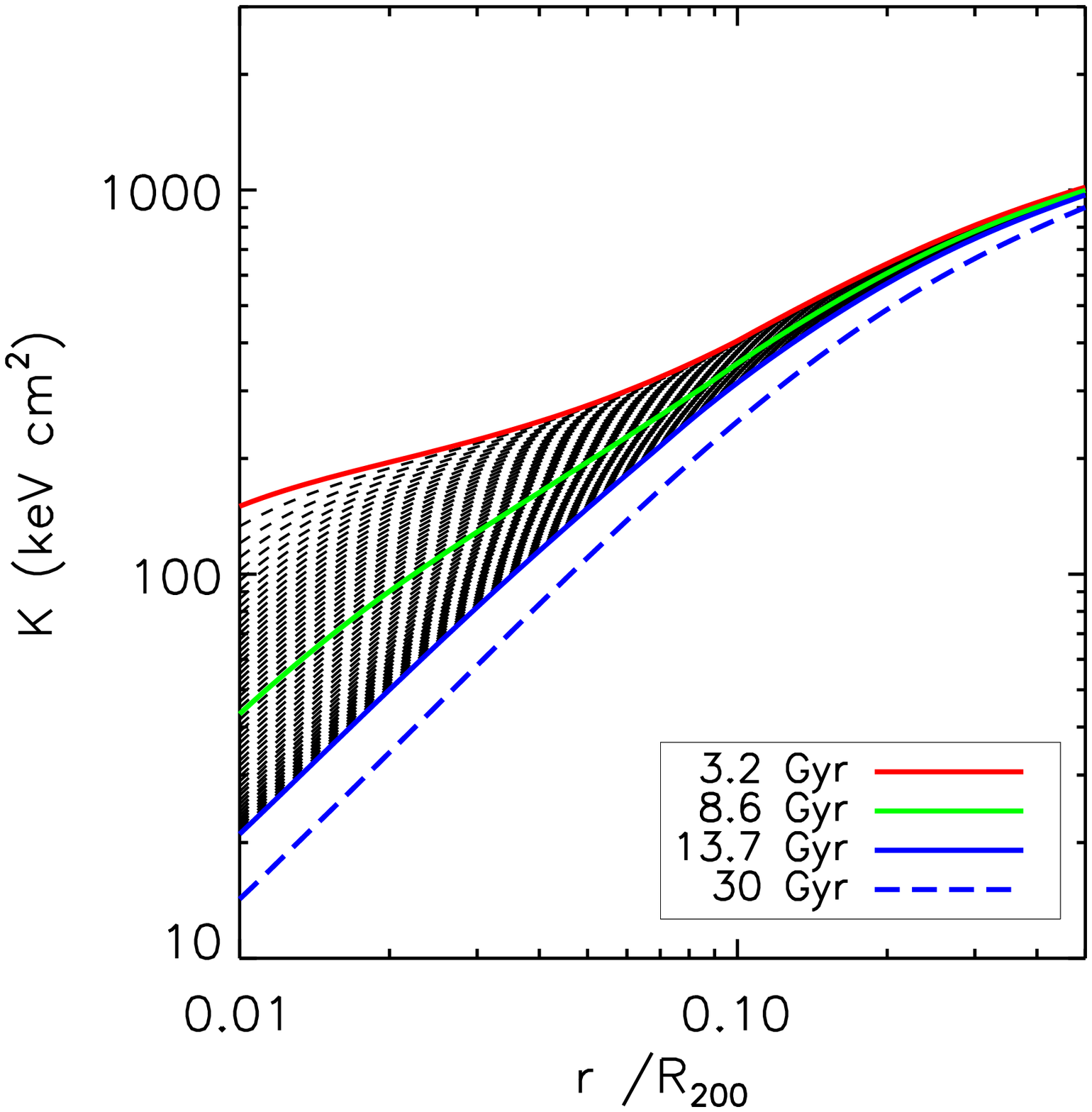,width=0.25\textwidth}
 \epsfig{figure=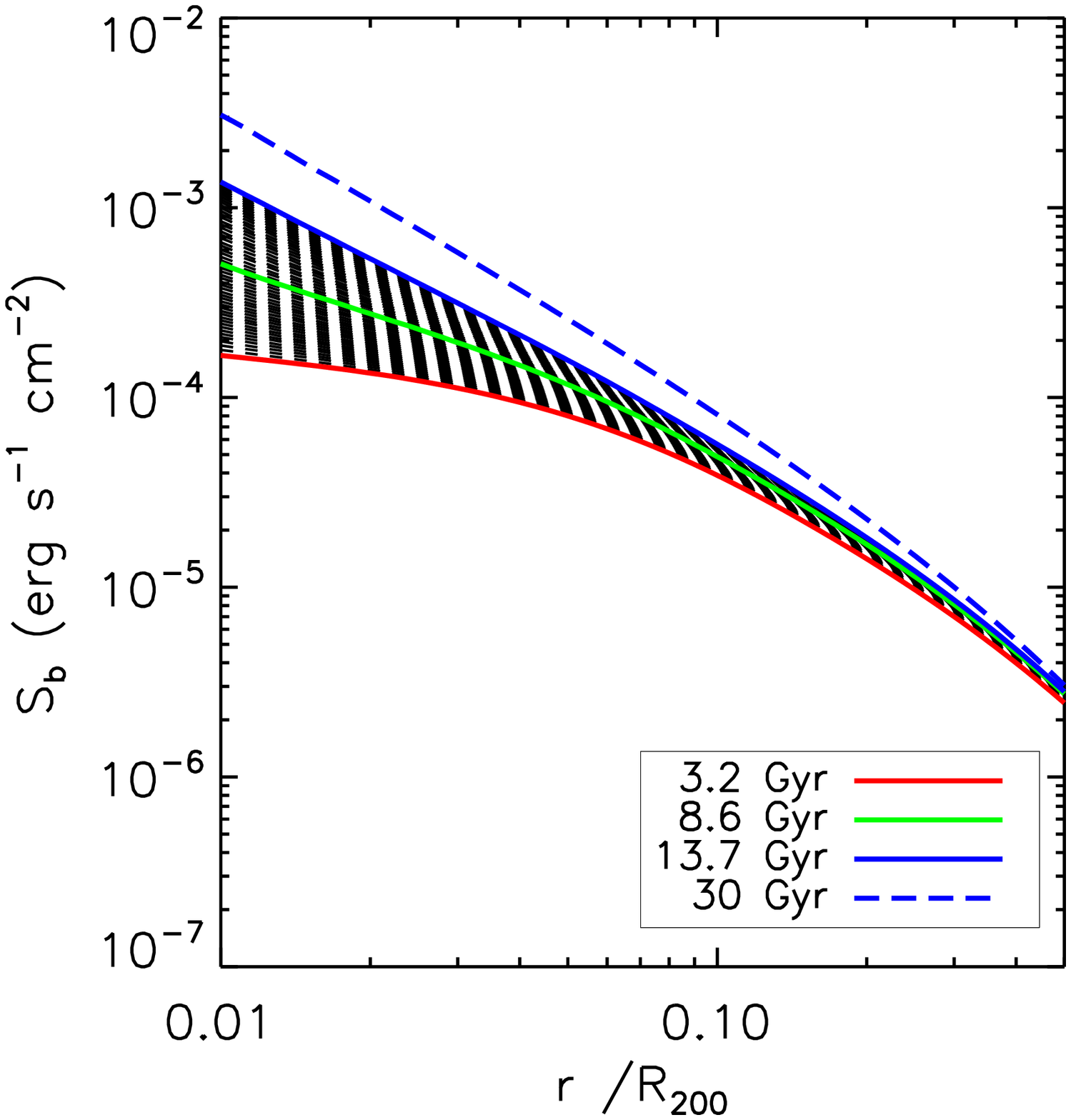,width=0.25\textwidth}
}
\hbox{
 \epsfig{figure=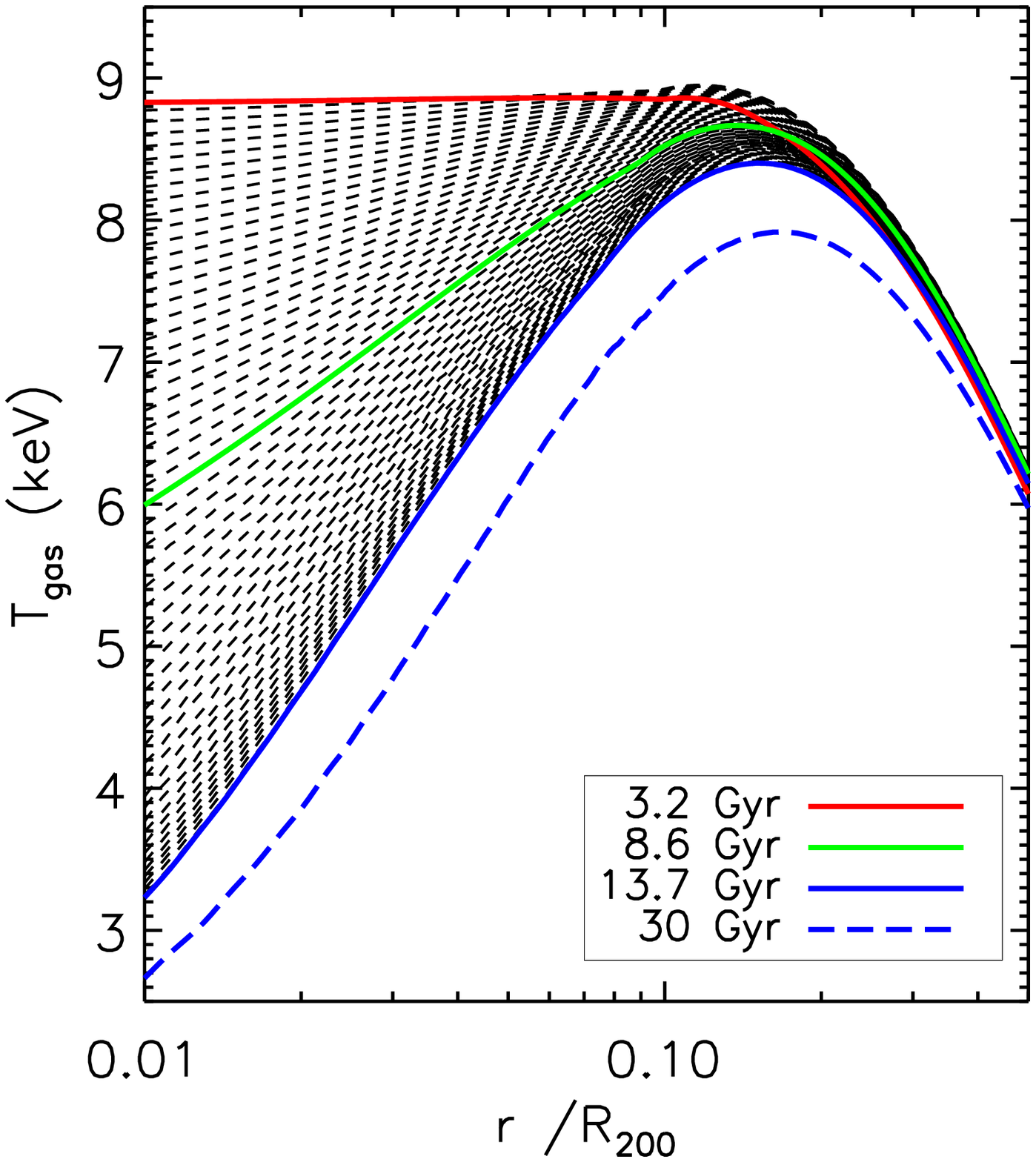,width=0.25\textwidth}
 \epsfig{figure=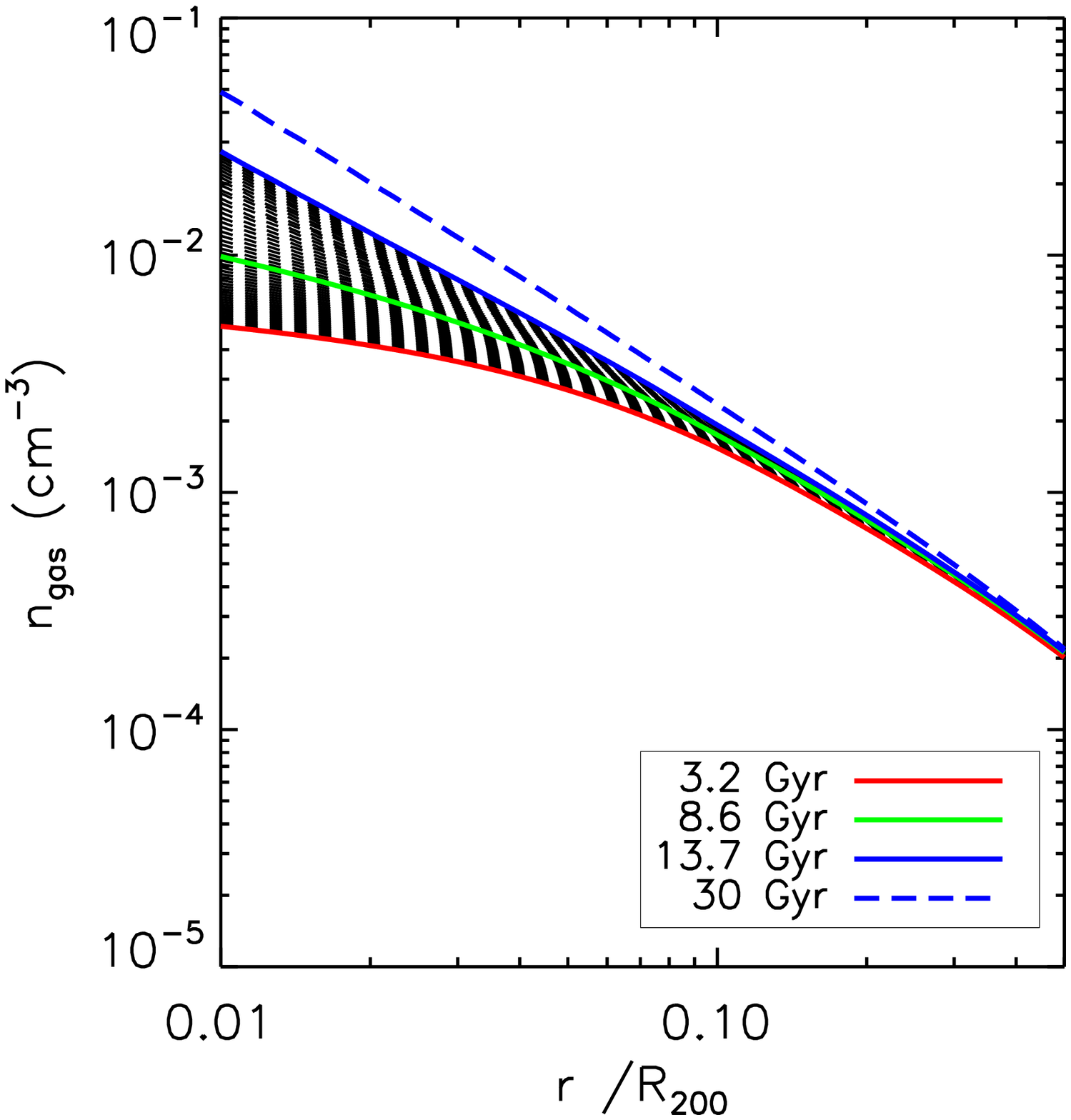,width=0.25\textwidth}
 \epsfig{figure=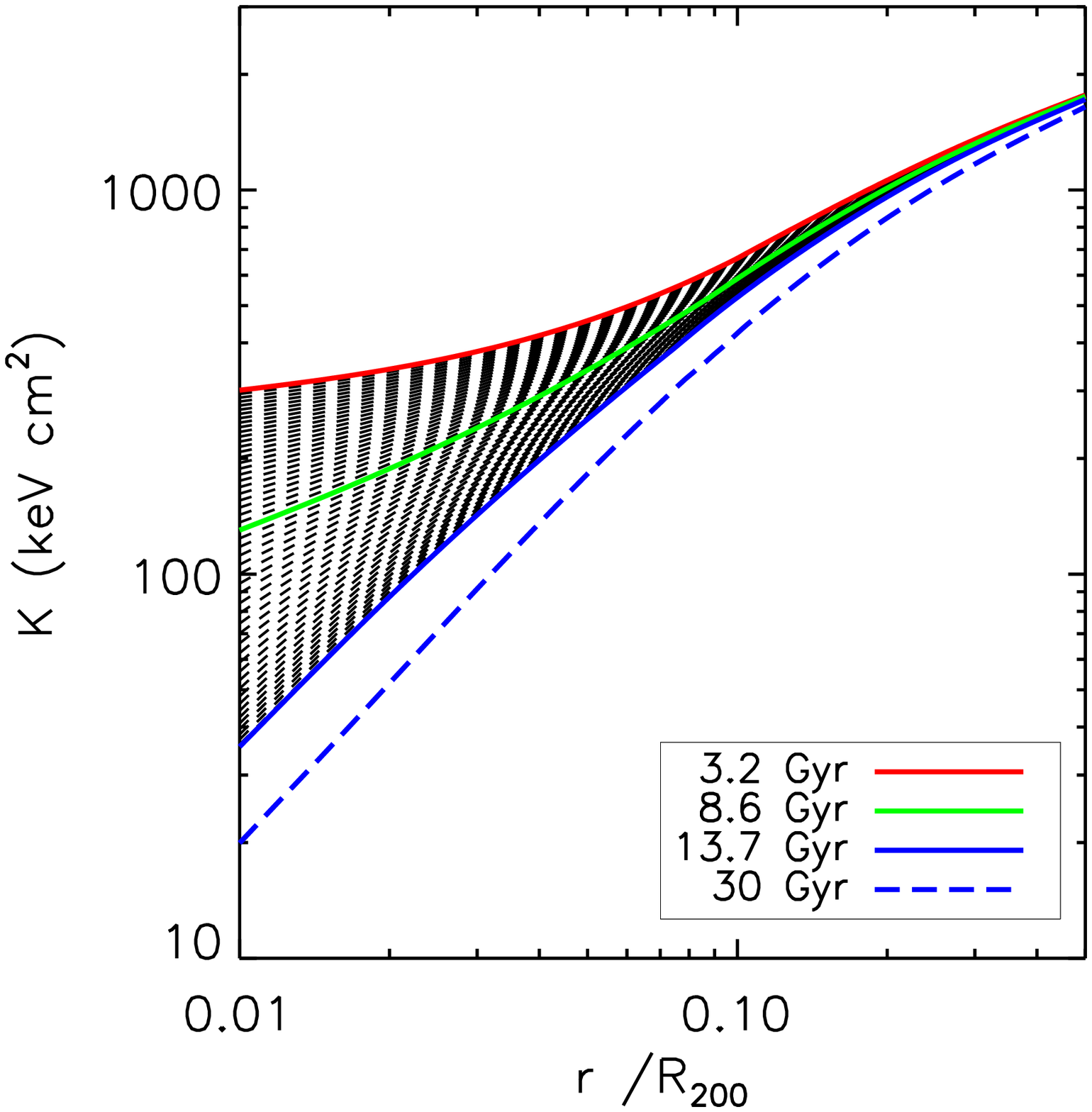,width=0.25\textwidth}
 \epsfig{figure=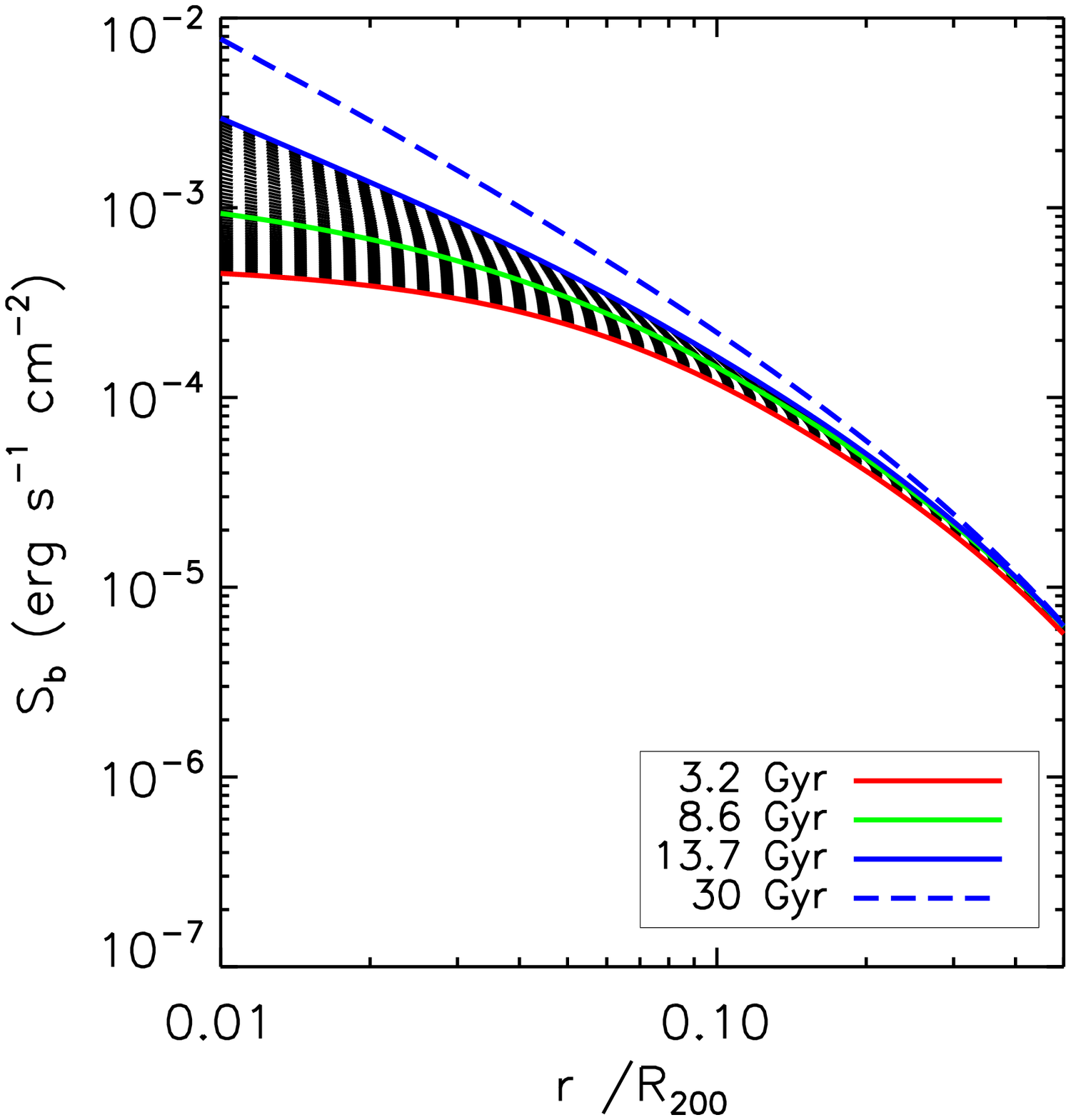,width=0.25\textwidth}
}
\hbox{
 \epsfig{figure=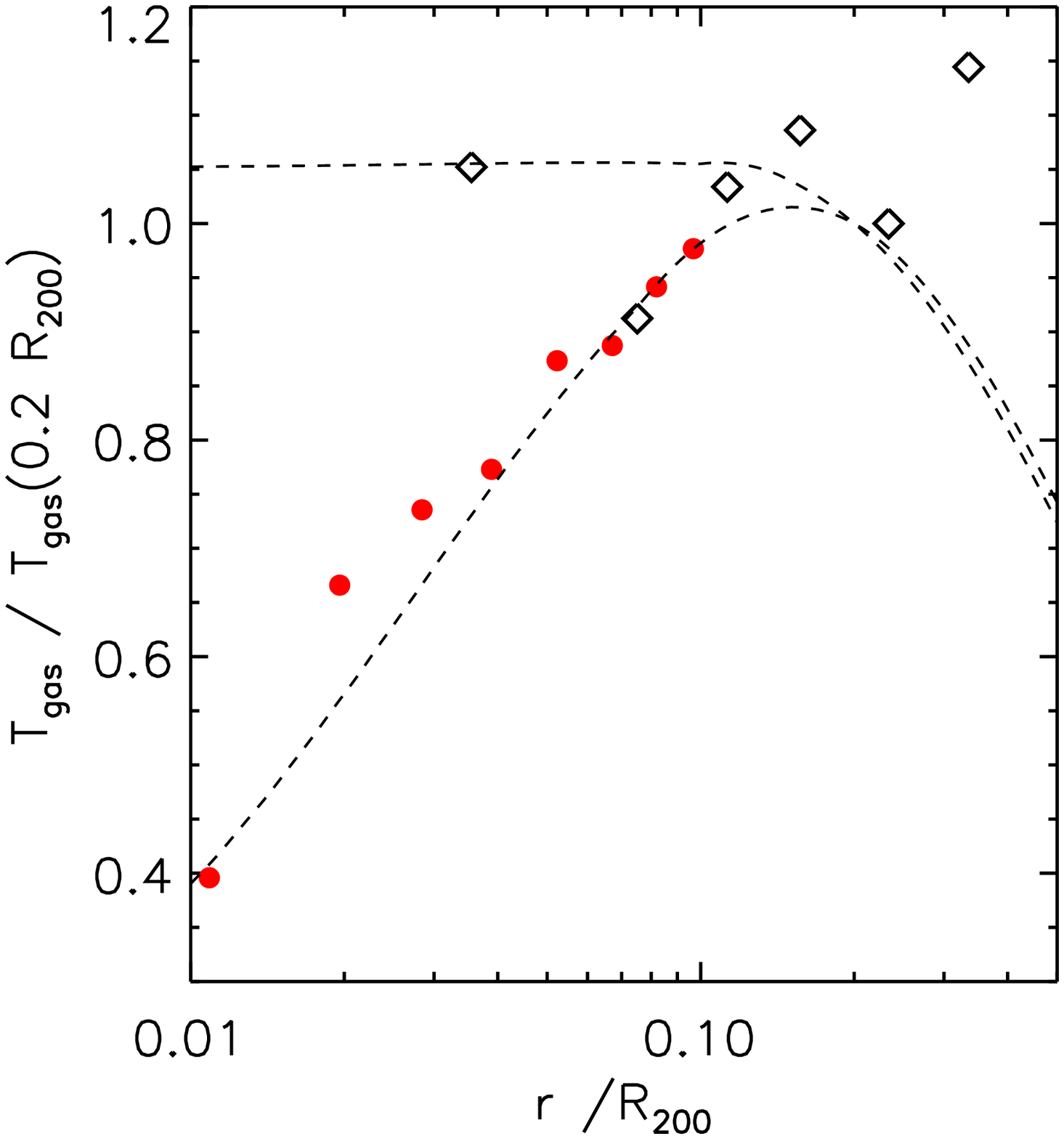,width=0.25\textwidth}
 \epsfig{figure=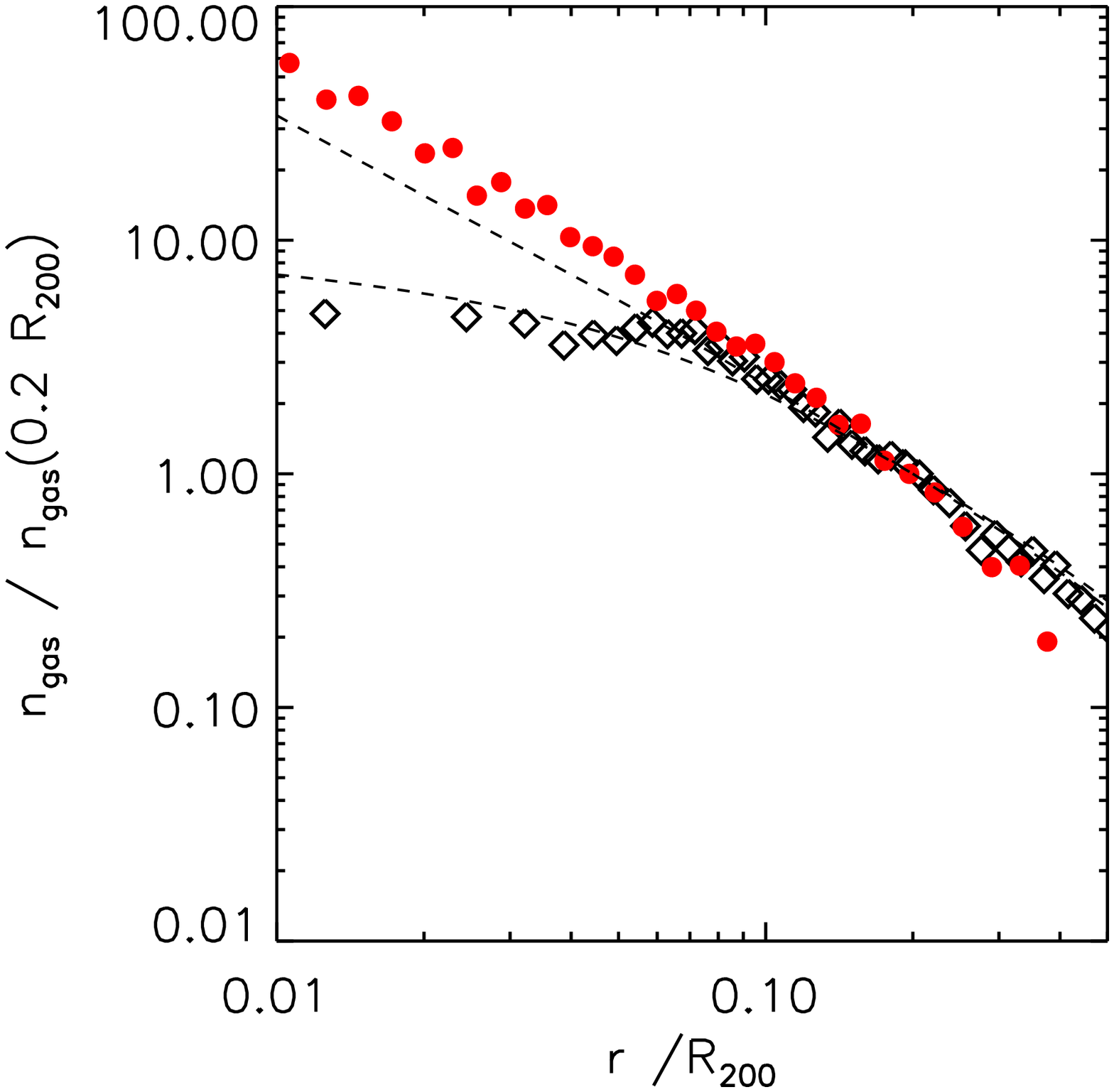,width=0.25\textwidth}
 \epsfig{figure=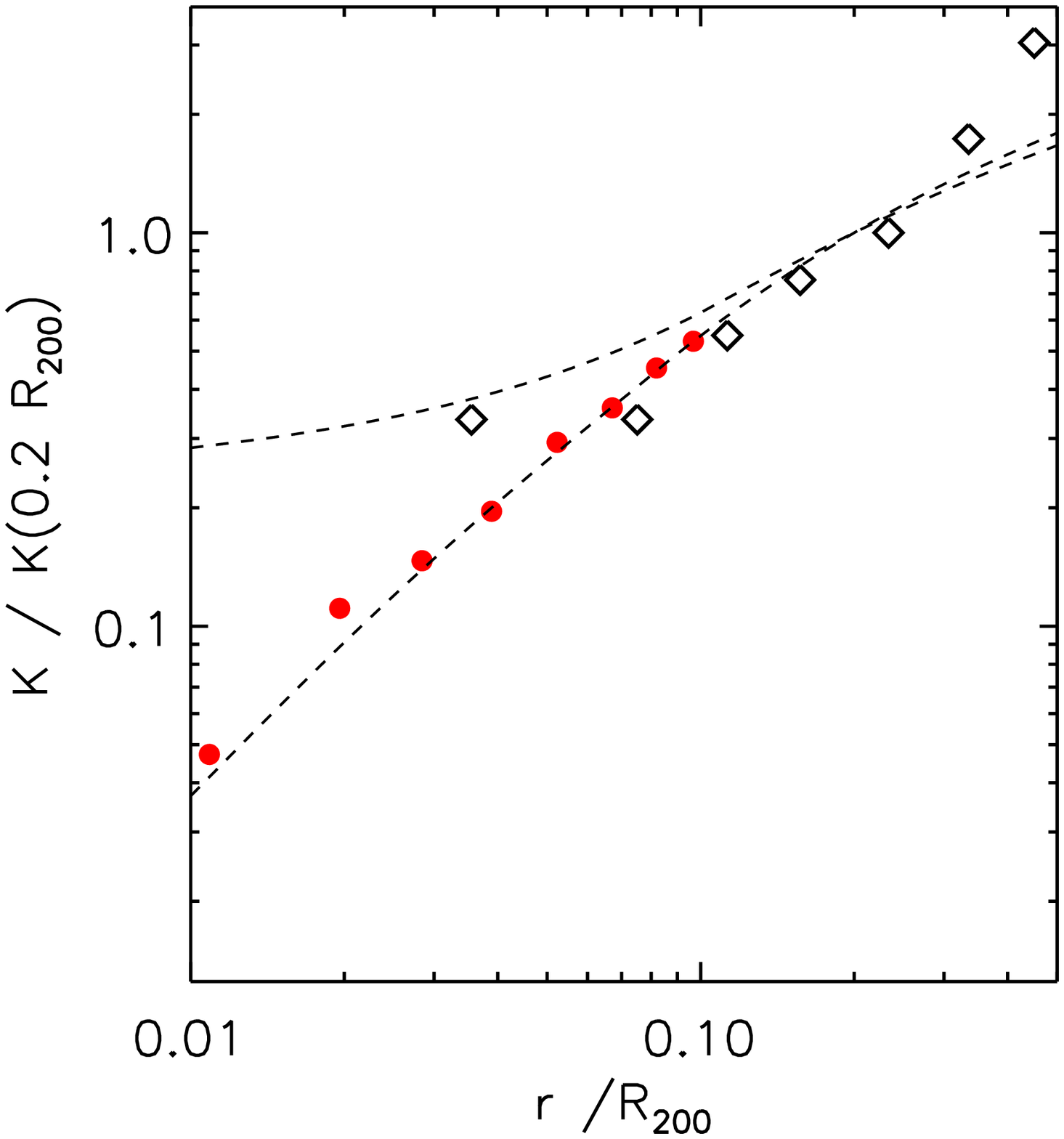,width=0.25\textwidth}
 \epsfig{figure=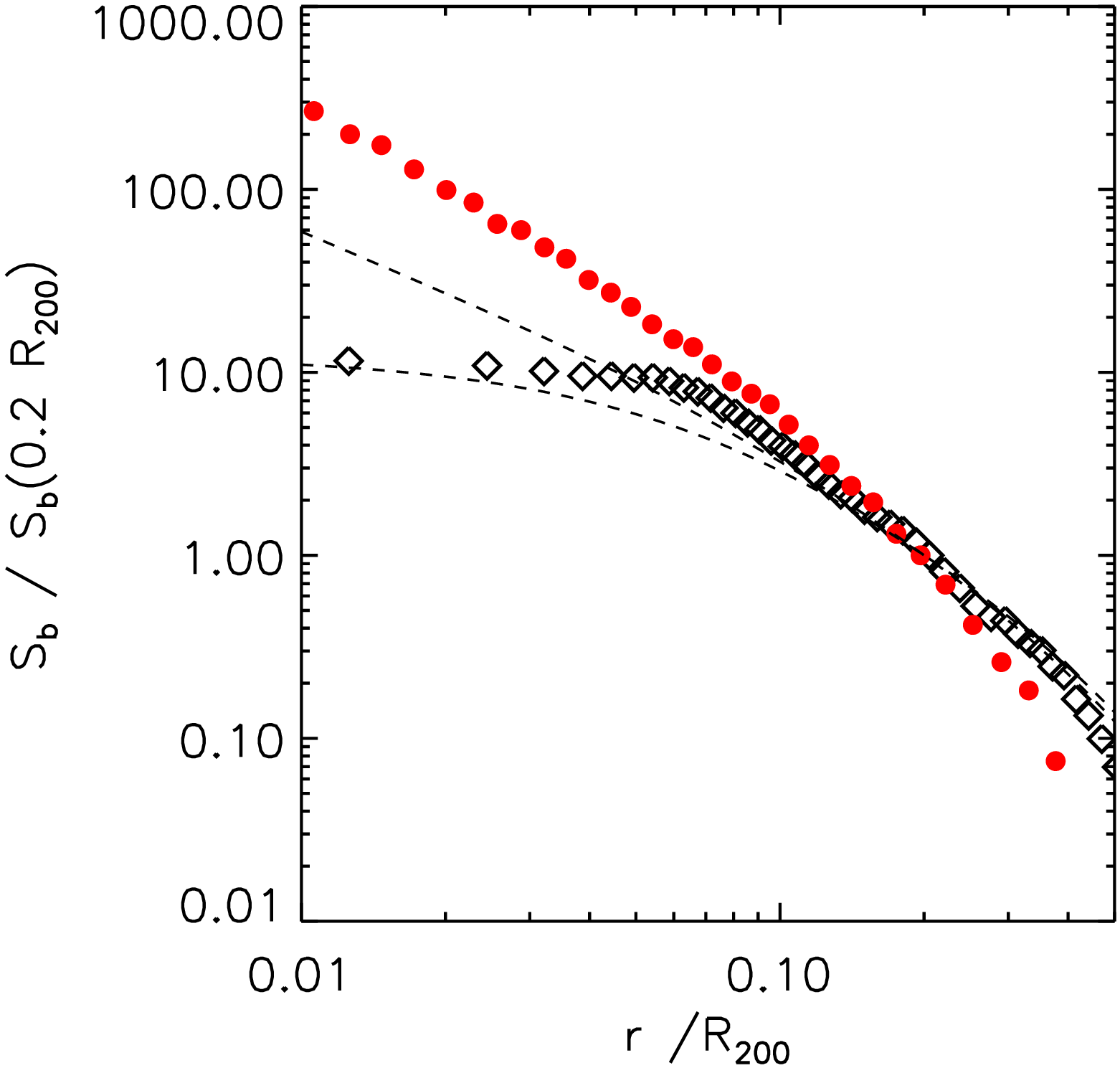,width=0.25\textwidth}
}
\caption{Evolution of the gas temperature, density, entropy and surface brightness 
profiles from redshift 2.1 ($t = 3.2$ Gyr) up to the present time ($t = 13.7$ Gyr) for the
``4 keV'' cluster ({\it upper panels}) and the ``8 keV'' cluster ({\it middle panels}).
For the latter case, we also show, for pedagogical reasons, the expected 
profiles at $t = 30$ Gyr, i.e. at about 16 Gyr in the future.
In the {\it lower panels}, we show the profiles renormalised at the
value at $0.2 R_{200}$ for a cool-core (A1795 and A1835; red dots) 
and a non-cool-core (A665; black diamonds) cluster.
The dashed lines show the simulated profiles at the beginning and end of the evolution. 
} \label{fig:prof} \end{figure*}

The paper is organized as follow: after the presentation of the 
numerical model adopted to describe the X-ray emitting plasma 
in hydrostatic equilibrium in a dark matter potential, we 
discuss in Section~3 the evolution in the observable properties
of a cooling core like the cooling radius, $R_{\rm cool}$,
the cooling time, $t_{\rm cool}$, the gas temperature, 
density, entropy and surface brightness profiles.
We present the models that reproduce the radial behaviour 
and the fitting parameters that
are more strongly correlated with the age of the structure,
being older structures the ones with the lower cooling time
at given radius.
A comparison with the observational constraints is also
discussed in Section~3.3. 
Finally, our results and discussion are summarized in Section~4.
Throughout this work we adopt the following cosmological 
parameters: $H_0=70$ km s$^{-1}$ Mpc$^{-1}$, $\Omega_{\rm m}=
1-\Omega_{\Lambda}=0.27$, that implies a present 
age of the Universe of 13.8 Gyr.

\begin{figure*}
\hbox{
 \epsfig{figure=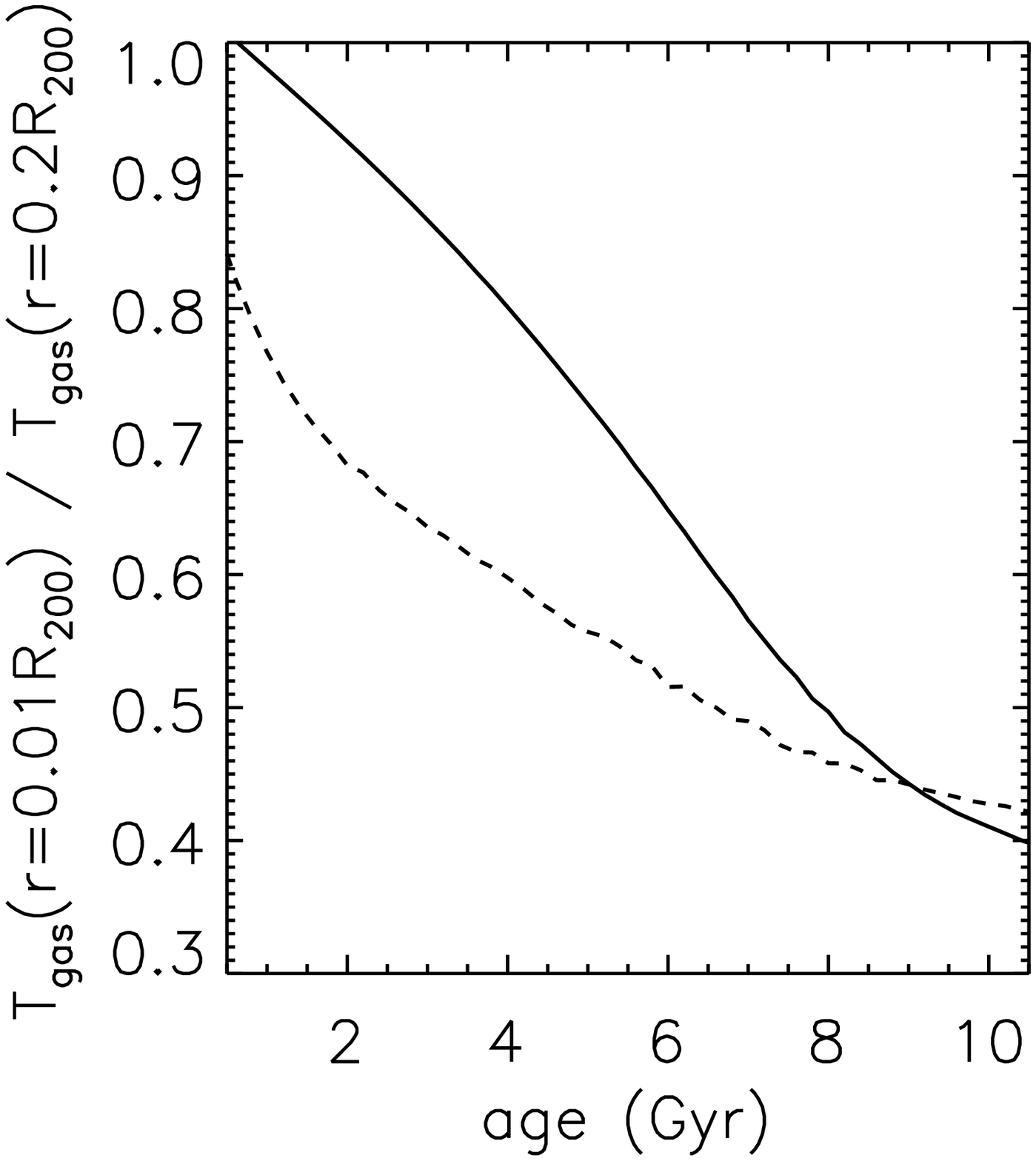,width=0.33\textwidth}
 \epsfig{figure=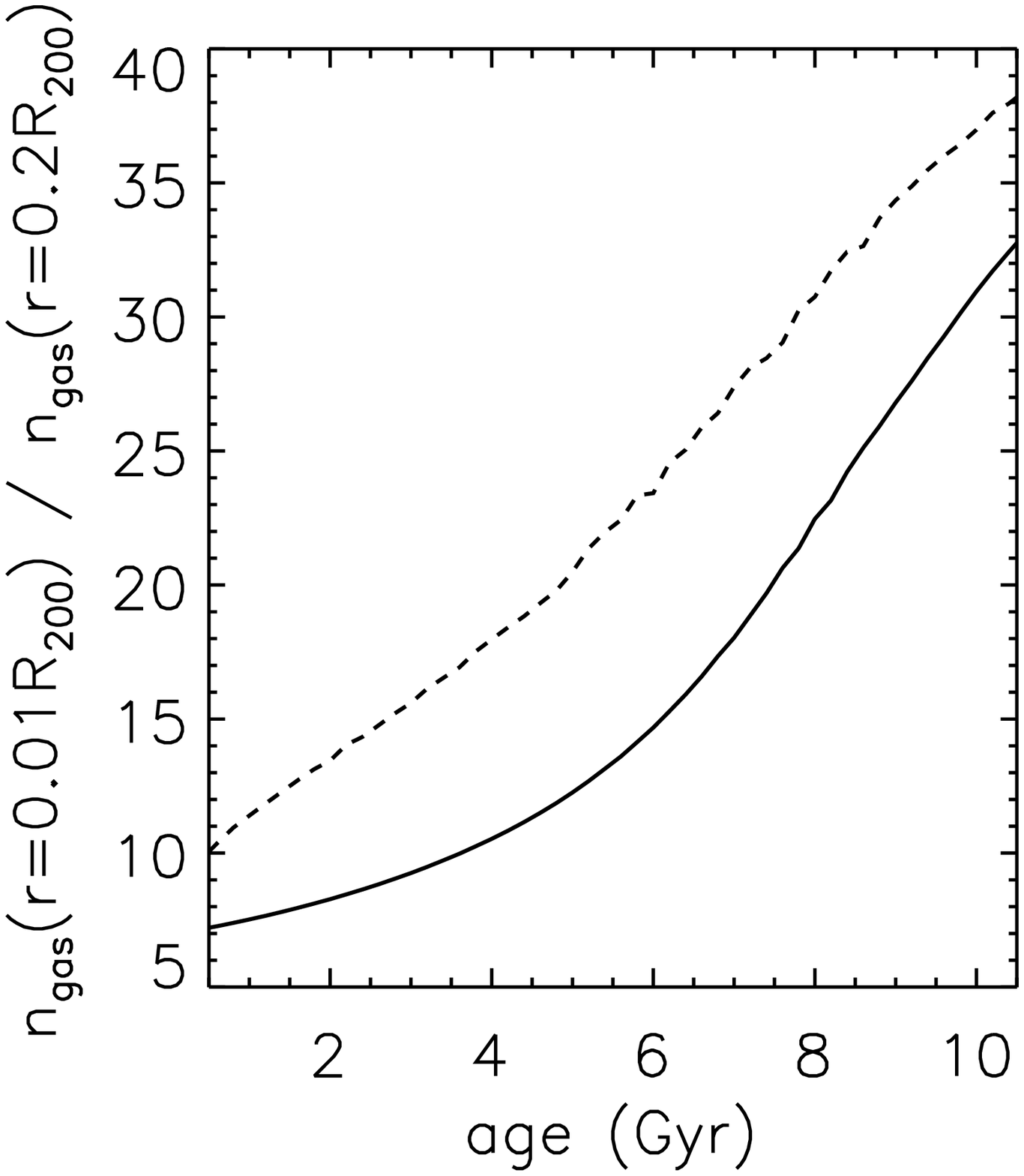,width=0.33\textwidth}
 \epsfig{figure=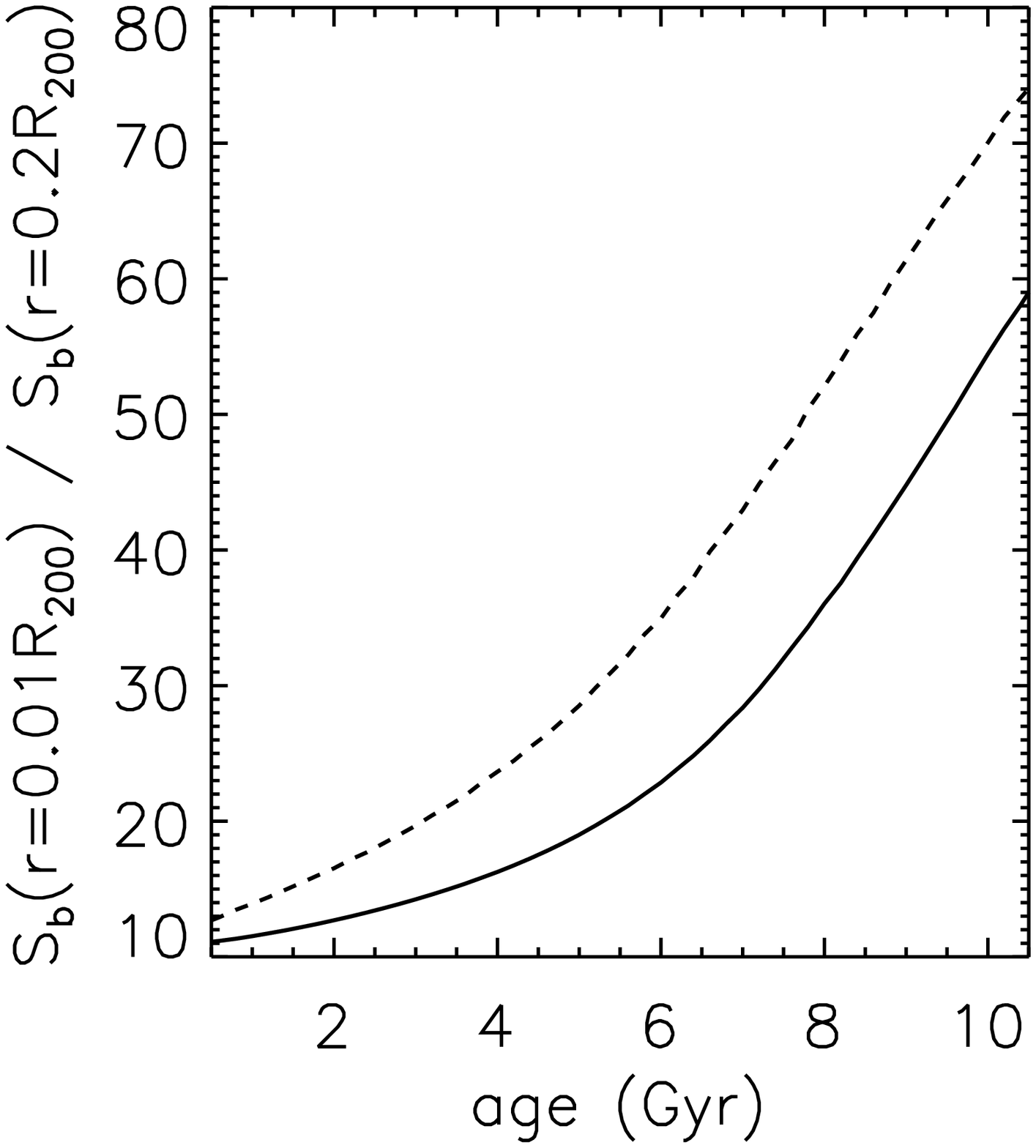,width=0.33\textwidth}
}
\hbox{
 \epsfig{figure=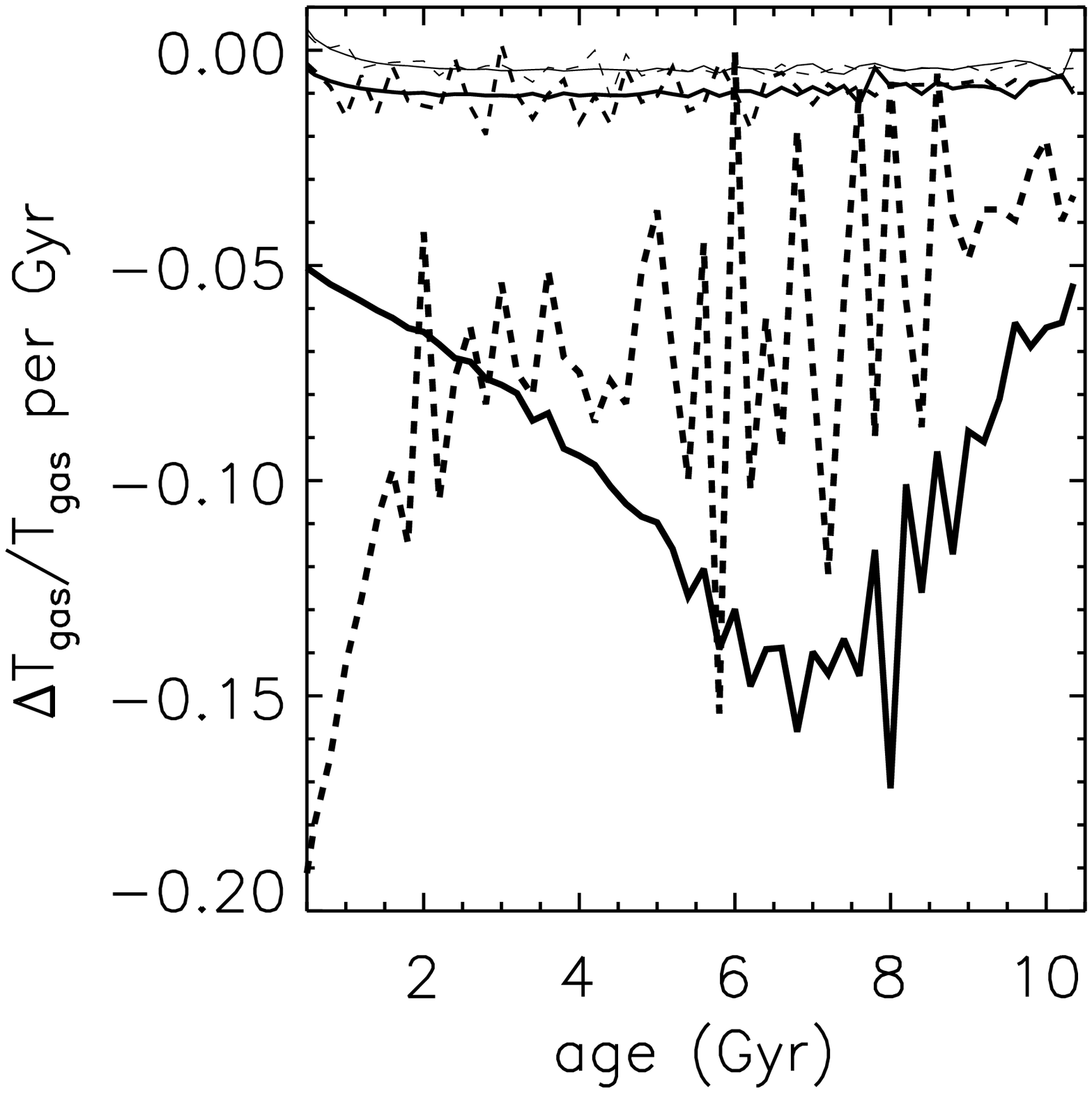,width=0.33\textwidth}
 \epsfig{figure=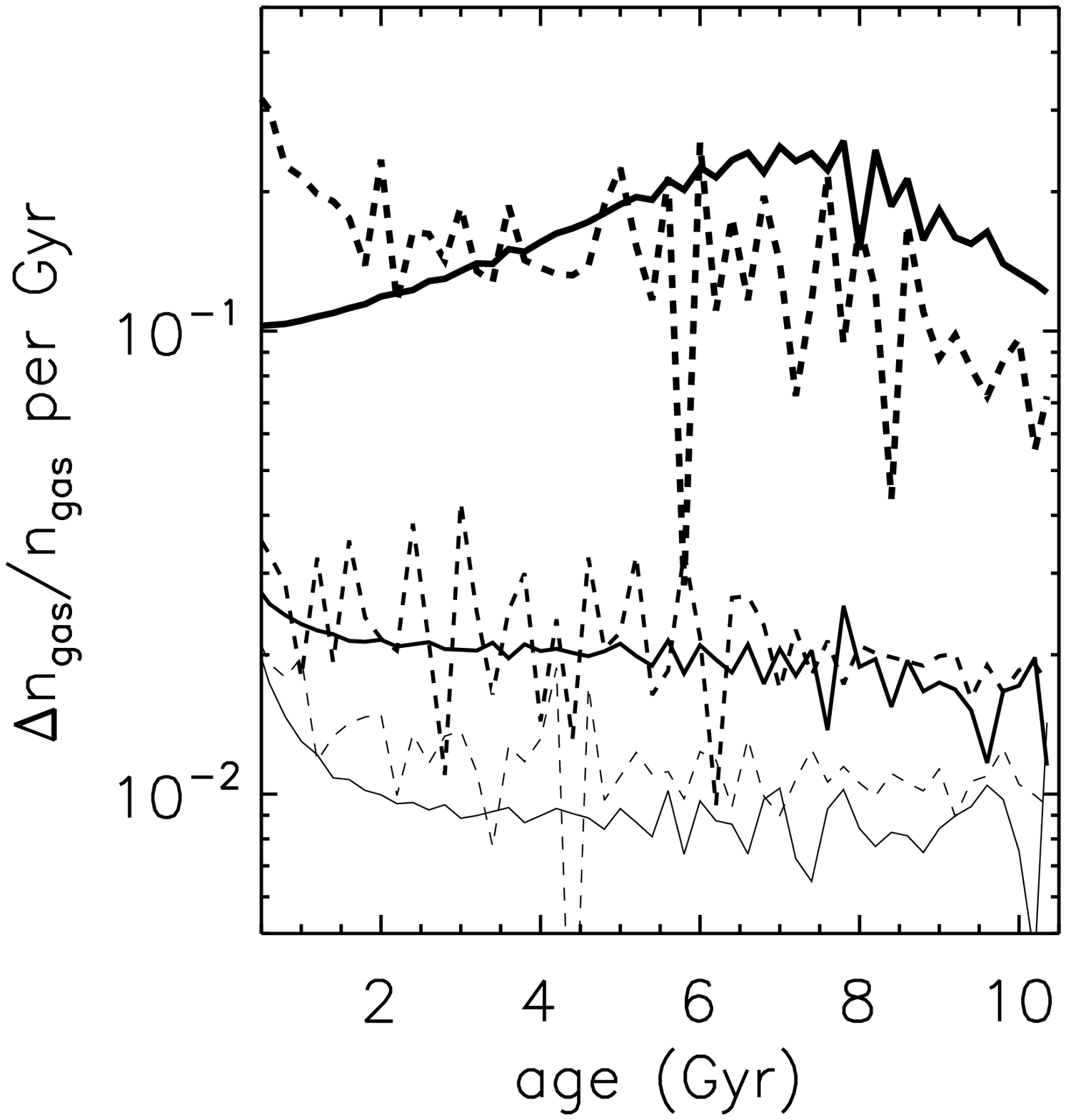,width=0.33\textwidth}
 \epsfig{figure=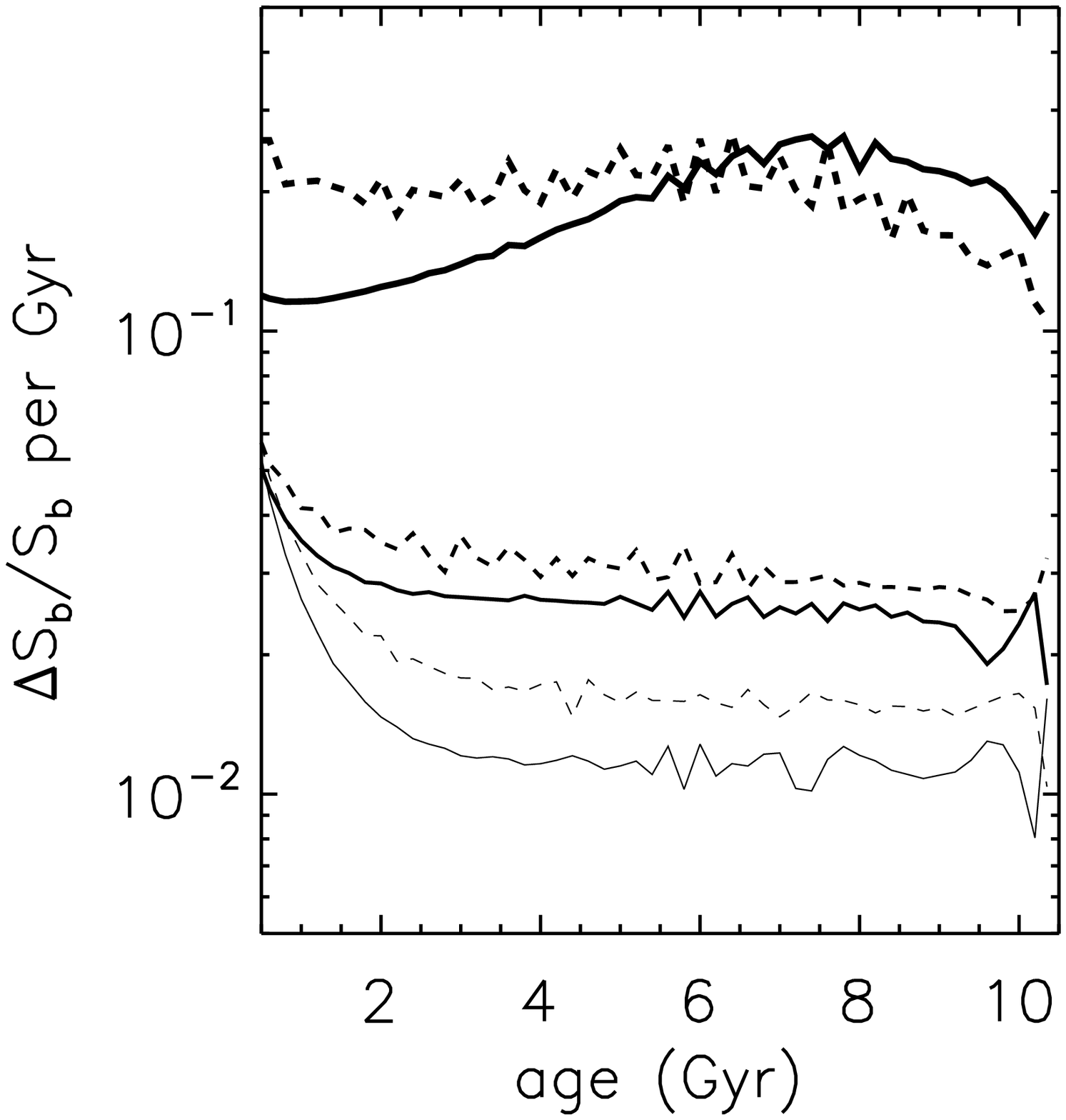,width=0.33\textwidth}
}
\caption{{\bf Upper panels} Ratios between the
gas temperature ({\it left}), density ({\it middle}) and surface brightness ({\it right})
estimated at $0.2 R_{200}$ and $0.01 R_{200}$ versus the age of the hot
({\it solid} line) and cool ({\it dashed} line) object.
{\bf Lower panels} Relative variation per Gyr as function of the structure's age
in the hot ({\it solid} line) and cold ({\it dashed} line) object.
The relative changes are estimated at $0.01, 0.1, 0.2 R_{200}$ from thickest
to thinnest lines, respectively.
} \label{fig:dq_age} \end{figure*}

\section{The numerical simulations}

The calculations presented below are based on the
time dependent 1D hydrodynamical
equations described in Brighenti \& Mathews (2002) with
the heating, conduction and convection terms set to zero. The cooling
function is modelled according to Sutherland \& Dopita (1993),
using a metallicity dependent fitting formula.
Here the abundance is set to 0.4 Z$_\odot$.

To solve the equations we use a modified version of the public code
ZEUS2D (Stone \& Norman 1992). We adopt spherical coordinates, and
the grid extends from $r=0$ to $r_{\rm max} \sim 10$ Mpc in 360 zones. 
The grid size is
$\Delta r = 0.5$ kpc at the center and slowly increases toward large radii.

Our simulations start with gas in hydrodynamical equilibrium 
in a potential well formed by a NFW dark matter halo (Navarro, Frenk 
\& White 1996) and a de Vaucouleurs profile representing the central galaxy.
Although not crucial, the presence of the galaxy and the associated
mass and energy
sources have second order effects on the ICM evolution. The reader is
referred to Brighenti \& Mathews (2002) for details about
stellar mass return and SNIa heating due to the central galaxy.
For the sake of simplicity, the gravitational potential is kept constant
in time. Neglecting the cosmic growth of the dark halo has some
influence in the inner structure of clusters, although halos grow
mostly inside-out and the gravitational potential in the core region
changes little with time during the smooth accretion phase between major
mergings (Zhao et al., 2003, Salvador-Sol\'e et al. 2007, Li et al. 2007).
The initial temperature profile follows the one described in Vikhlinin et al.
(2006a) for $r\ge 0.15\,R_{500}$ (where $R_{500}$ is defined below): 
$$T(r) = 1.216 T_{\rm x}\, {(x/0.045)^{1.9} + 0.45\over (x/0.045)^{1.9}+1}
{1 \over [1 + (x/0.6)^2]^{0.45} }$$
while is isothermal $T(r)=T(0.15\,R_{500})$ for $r< 0.15\,R_{500}$.
Here $T_{\rm x}$ is the global (spectroscopic)
temperature, calculated according to the observational
mass-temperature relation by Arnaud et al. (2005) 
(see also Shimizu et al. 2003). 
The adopted initial temperature profile should represent a reasonable
initial condition to study the radiative evolution of the cluster cores.
The outer region ($r\ge 0.15\, R_{500}$) is not greatly influenced
by radiative cooling and thus we decided to set it to agree with
current observations.
In section 3.3 we briefly discuss how the results depend on the choice
of the initial conditions.

We present results for two cluster models.
The ``hot'', massive cluster has a virial dark matter mass 
$M_{\rm DM,hot} = 1.3 \times 10^{15}$
M$_\odot$, while the ``cold'' object has  $M_{\rm DM,cold} = 
4.1 \times 10^{14}$ M$_\odot$. The concentration
of the dark matter halo is calculated by fitting the $c-M$ relation
given in Bullock et al. (2001): $c=8.35 \, (M_{\rm DM}/10^{14} 
\;{\rm M}_\odot)^{-0.911}$.
The central galaxy is the same for the two
models and has a stellar density profile following the approximation
for a deprojected de Vaucouleurs law,
given by Mellier \& Mathez (1987). The total mass of the
galaxy is assumed to be $M_* = 5.8 \times 10^{11}$ M$_\odot$ and the
effective radius is $R_e = 8.5$ kpc. These numbers are typical for
giant elliptical galaxies. The initial gas temperature are $T_0=7.67$ keV
and $T_0=4.0$ keV for the hot cluster and the cold cluster respectively.
The initial central gas density is set by the requirement that the baryon
fraction at the virial radius is similar to the cosmic one, i.e. we require 
$f_{\rm b}(R_{\rm vir})=0.16$ (e.g. Spergel et al. 2007). 
This results in $\rho_0=2.26 \times 10^{-26}$ g cm$^{-3}$
for the cold cluster and $\rho_0=1.82 \times 10^{-26}$ g cm$^{-3}$ 
for the hot cluster.
The central cooling time, defined as $t_{\rm cool} = 2.5 k\rho T/
\mu m_p/n_{\rm e}n_{\rm H}\Lambda(T) $, is $\sim 3.4$ Gyr (cold cluster)
and $\sim 8.5$ Gyr (hot cluster).

In order to better compare the modelled temperature with observations
we present results for both the emission weighted temperature
$T_{ew}(r)$
and the ``spectroscopic-like'' temperature $T_{sp}(r)$ defined in 
Mazzotta et al. (2004). $T_{ew}$ is calculated in the usual way as
the line of sight integral $T_{ew}=\int{T \epsilon dl}/\int{\epsilon dl}$, 
where $\epsilon$ is the (bolometric) gas 
emissivity, while $T_{sp}=\int{T (n^2 T^{-0.75}) dl}/\int{n^2 T^{-0.75} dl}$.
According to Mazzotta et al. (2004) $T_{sp}$ is a good approximation
to the temperature measured with ${\it Chandra}$ and ${\it XMM}$ when
the thermal structure of the ICM is complex. By performing the integral
along the line of sight at any radius, we calculate a projected
temperature profile which can be directly contrasted with the observed ones.
The global temperature in a given region (i.e. $0<r<R_{500}$) is then 
calculated with a similar weighted average: $T_{sp}(0-R_{500}) = 
\int_0^{R_{500}} T_{sp}(r) (n^2 T^{-0.75}) 4\pi r^2 dr / 
\int_0^{R_{500}} n^2 T^{-0.75} 4\pi r^2 dr$. 

In the following we will normalize the radial distances by $R_{200}$ or
$R_{500}$, the radii at which the overdensity is 200 or 500 respectively.
For our two models we have: $R_{200}=1.41$ Mpc, $R_{500}=0.93$ Mpc (cold cluster)
and $R_{200}=2.05$ Mpc, $R_{500}=1.34$ Mpc (hot cluster).

\section{The evolution of the radial profiles}

The calculations described here are similar to the unheated flow
presented in Brighenti \& Mathews (2003), where the reader may find
a description of the thermal evolution of the flow. Recently,
Guo \& Oh (2007) have also discussed simulations of the ICM in a $T\sim 4$
keV cluster. The results of our calculations are in qualitative agreement
with those in the two aforementioned papers.
As the gas cools in the central region, it slowly flows inward to 
approximately preserve equilibrium. A positive temperature gradient
develops and the density profile steepens.
At the end of the simulations the temperature profile agrees nicely
with typical observed cool core clusters (Figure 1), while the computed
density is somewhat flatter. This can be understood with the long
initial central cooling time (see above), which makes the ``radiative age''
of our models rather young, even after $\sim 10$ Gyr.
After about one cooling time the gas in the centre cools to very low
temperature, this is the onset of the so-called cooling catastrophe.
The mass cooling rate grows with time, reaching an approximate steady-state
after few cooling times (see Brighenti \& Mathews 2003, 2006). In the
calculations presented here the mass cooling rates are still increasing
at the end of the simulations, 
due to the rather long initial cooling times.
At $t=13.7$ Gyr we find
$\dot M_{\rm cool}\approx 70$ $(40)$ $M_\odot$/yr
for the hot (cold) cluster. A total mass of
$\sim 1.1 \times 10^{11}$ $M_\odot$ and $\sim 1.3\times 10^{11} M_\odot$
is cooled below X-ray temperature in the hot and cold cluster
respectively.
This is at odd with observations, which show tight constraints
on the cooling rate (e.g. Peterson \& Fabian 2006) -- a blatant manifestation
of the cooling flow problem. We do not attempt to solve this problem here
(see McNamara \& Nulsen 2007 for a recent 
review on this subject). Rather,
we are interested in the formation of the cool core following the 
assembly of the cluster or the last major merging. The way in which
the heating necessary to halt radiative cooling influences the density
and temperature profiles is not well known. Here we take the position
that the average thermal structure of the ICM is not greatly affected
by the heating process. This appears justified by the reasonable agreement
shown in Figure 1 below between observed and simulated, purely radiative
temperature profile (see Brighenti \& Mathews 2006 and Guo \& Oh 2007 for
models in which the variables profiles are not
significantly modified by jet heating). 
Following this assumption we neglect the gravity from the unrealistic 
accumulation of cold gas in the centre of our models, which would
generate a strong temperature peak (Brighenti \& Mathews 2000).

The radial profiles of the gas temperature, density and surface brightness
are obtained at temporal steps of 0.2 Gyr starting from the cosmic time
of 3 Gyr, corresponding to $z=2.11$ for the assumed cosmology,
and reaching $z=0$ after 10.7 Gyr.
These profiles are plotted in Fig.~\ref{fig:prof}.
When we refer to the age of the structure, we mean
the cosmic time at which the physical quantity is observed
{\it minus} the cosmic time at which the structure started
its evolution, i.e. 3 Gyr.
In the same figure, we present the profiles normalized at the
values observed at $0.2 R_{200}$ in a cool-core (A1835 
from Morandi \& Ettori 2007 and A1795, for the inner regions,
from Ettori et al. 2001)
and a non-cool-core (A665, Morandi \& Ettori 2007) massive cluster. 
The profiles follow the predicted behaviour in particular of the
gas density and surface brightness distribution. Larger deviations
are observed in the temperature profile that depend strongly on the
reference global value adopted to normalize the profile.

We present in Fig.~\ref{fig:dq_age} the relative variation
of the temperature, density and surface brightness as function
of the age of the objects, both as ratios between
the quantities estimated at fixed fractions of $R_{200}$
(i.e. $0.01 R_{200}$, that lies well within the cluster core, and
$0.2 R_{200}$, radius at which the cooling is not effective anymore)
and as relative changes per Gyr at $0.01, 0.1, 0.2 R_{200}$.
The expected behaviour is confirmed for all the quantities under exam
with the added value that we are now in condition to quantify the magnitude
of the variations. 

The temperature decreases rapidly at first, with relative changes between 5 and
20 per cent per Gyr when measured at $0.01 R_{200}$. After $\sim 7$
Gyr the temperature profile of the two objects reaches a 
quasi-steady state. This results in a ratio $T (r=0.01R_{200}) / T (r=0.2R_{200})$ 
equals to $\sim0.4$ after 10 Gyr of pure cooling flow evolution.

The gas density rises at a rate of about 15-20 per cent per Gyr
when measured at $0.01 R_{200}$ and of 1 per cent when
the increase is evaluated at $0.2 R_{200}$.
After 10 Gyr, the ratio between these values is about 35,
a factor $\sim 4-5$ larger than the ratio measured at the beginning
of the formation of the cooling core.

The surface brightness, roughly proportional to the integral along the
line of sight of the squared density,
amplifies the variations observed in the density
profile and is observed to increase with a mean rate of 16 and 1 per cent
per Gyr at $0.01$ and $0.2 R_{200}$, respectively, 
with a ratio that rises from 10 to 60--70 at the present time, both
in cool and hot systems.

\begin{figure*}
\hbox{
 \epsfig{figure=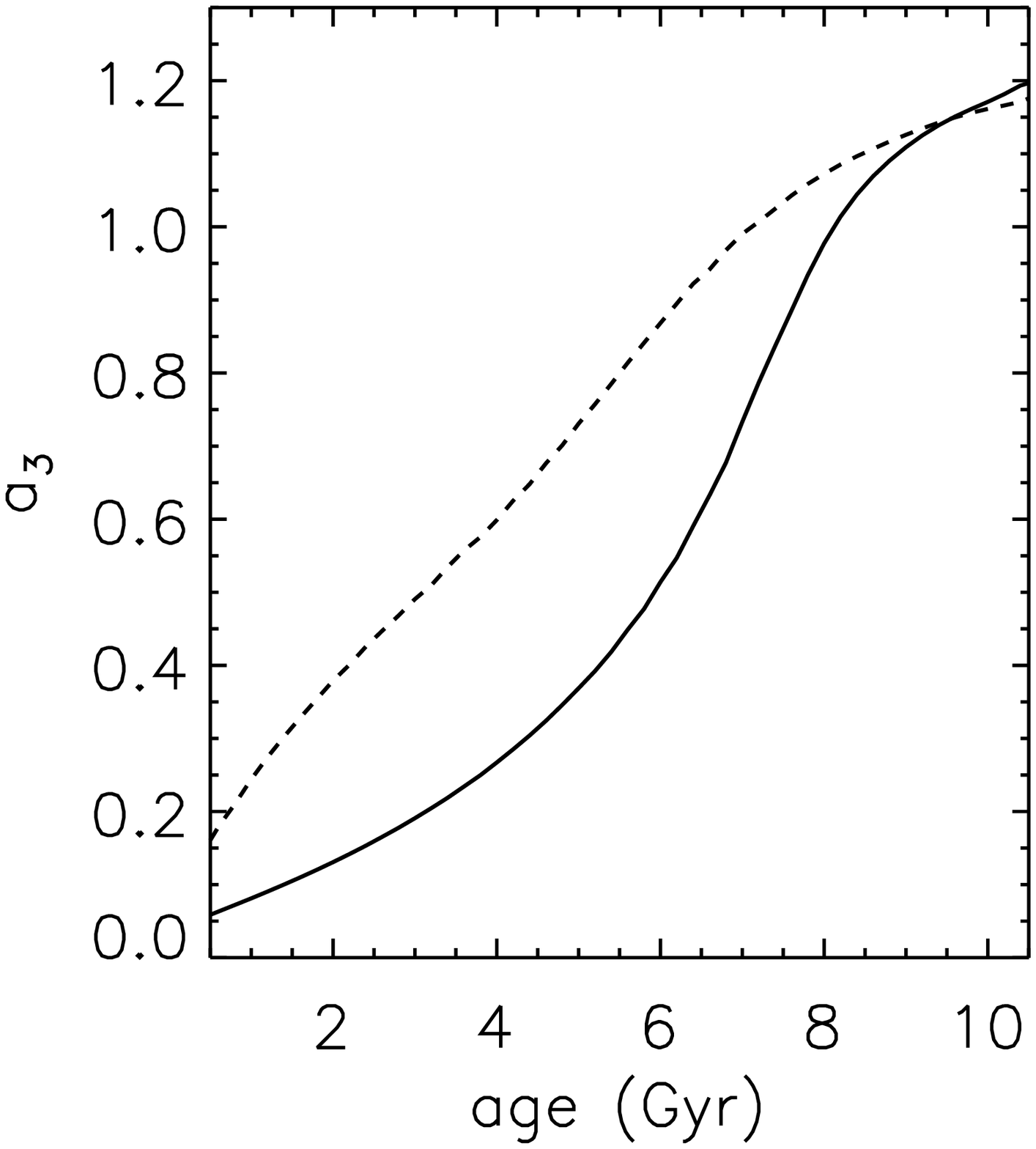,width=0.33\textwidth}
 \epsfig{figure=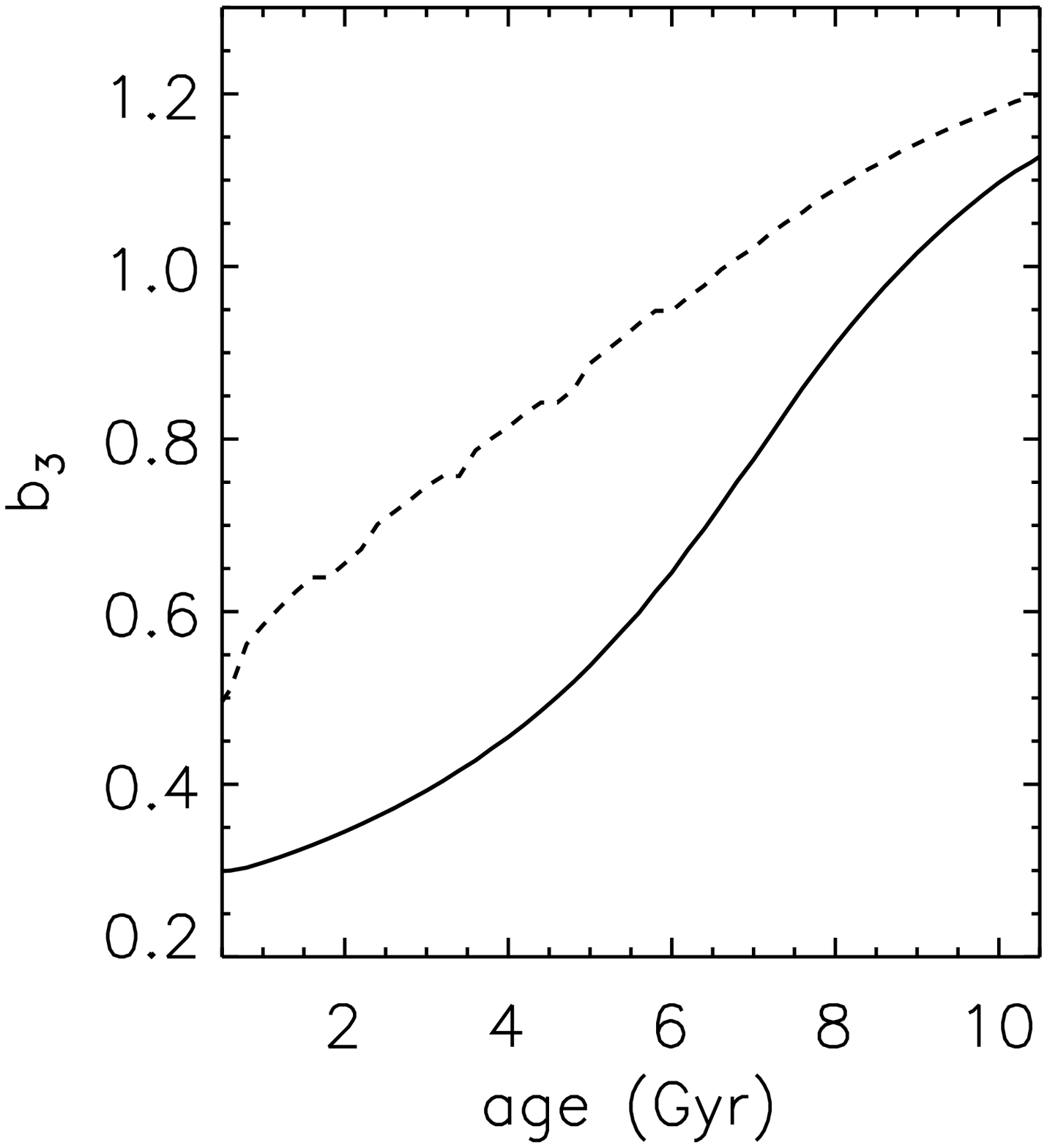,width=0.33\textwidth}
 \epsfig{figure=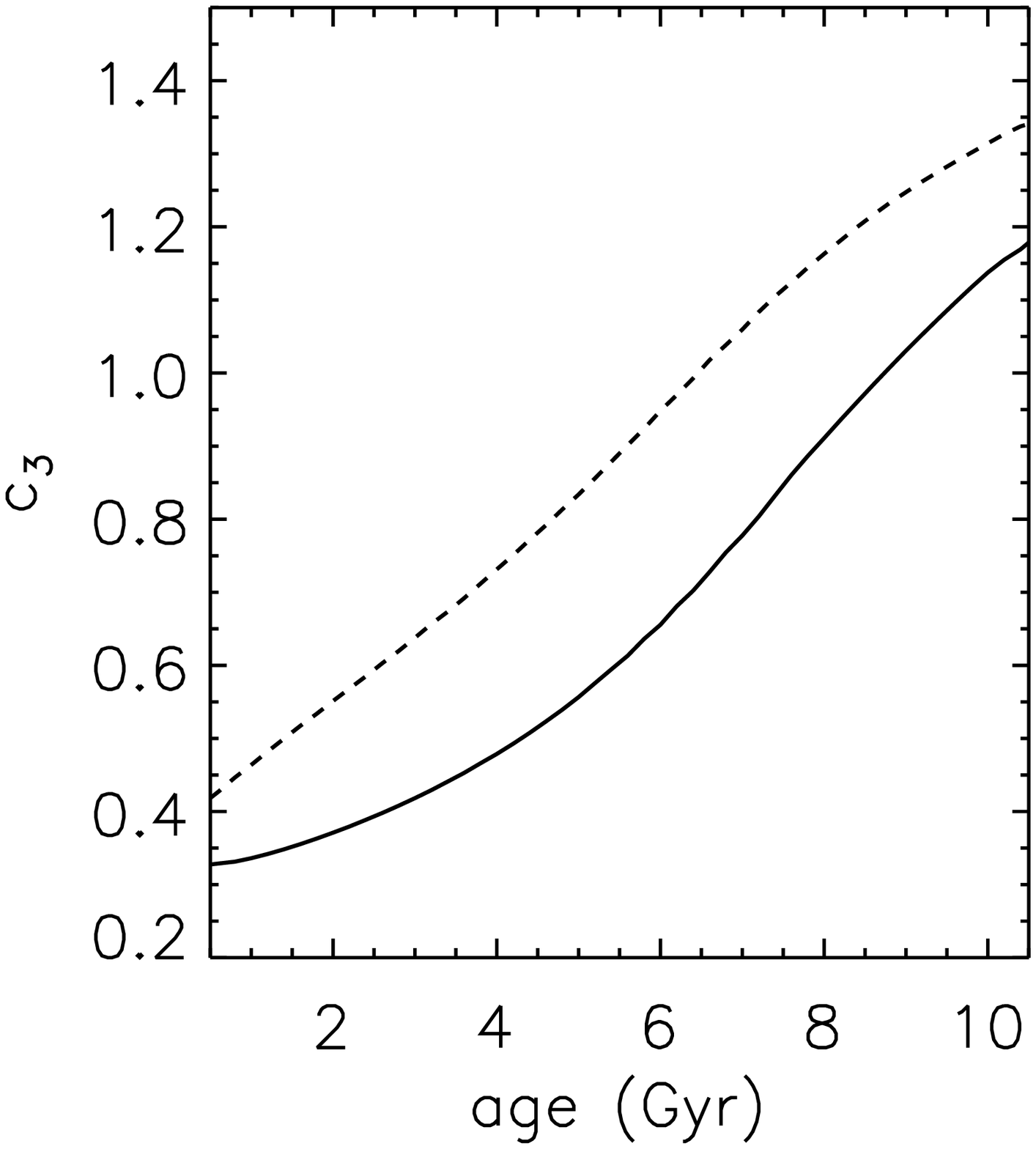,width=0.33\textwidth}
}
\hbox{
 \epsfig{figure=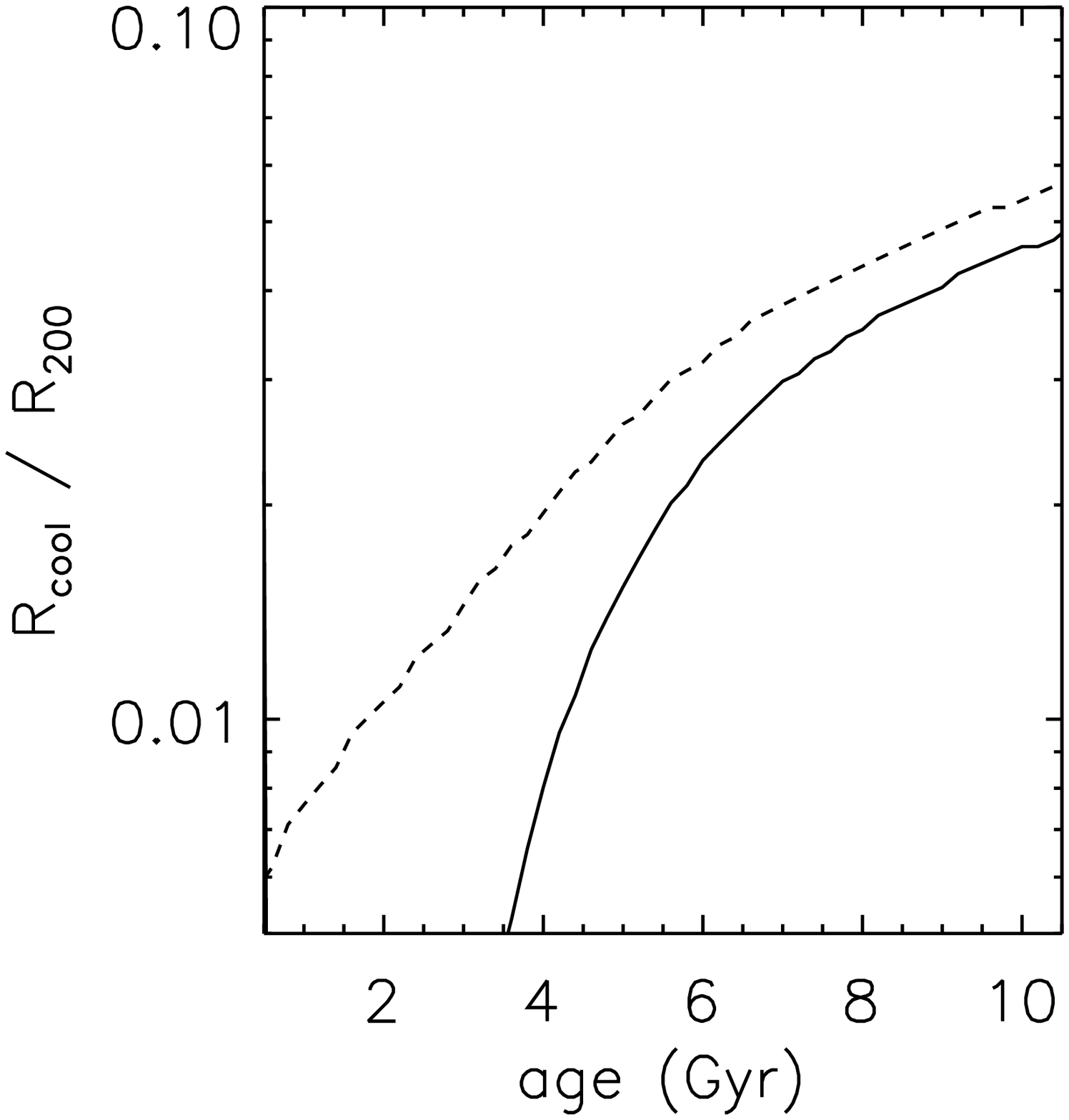,width=0.33\textwidth}
 \epsfig{figure=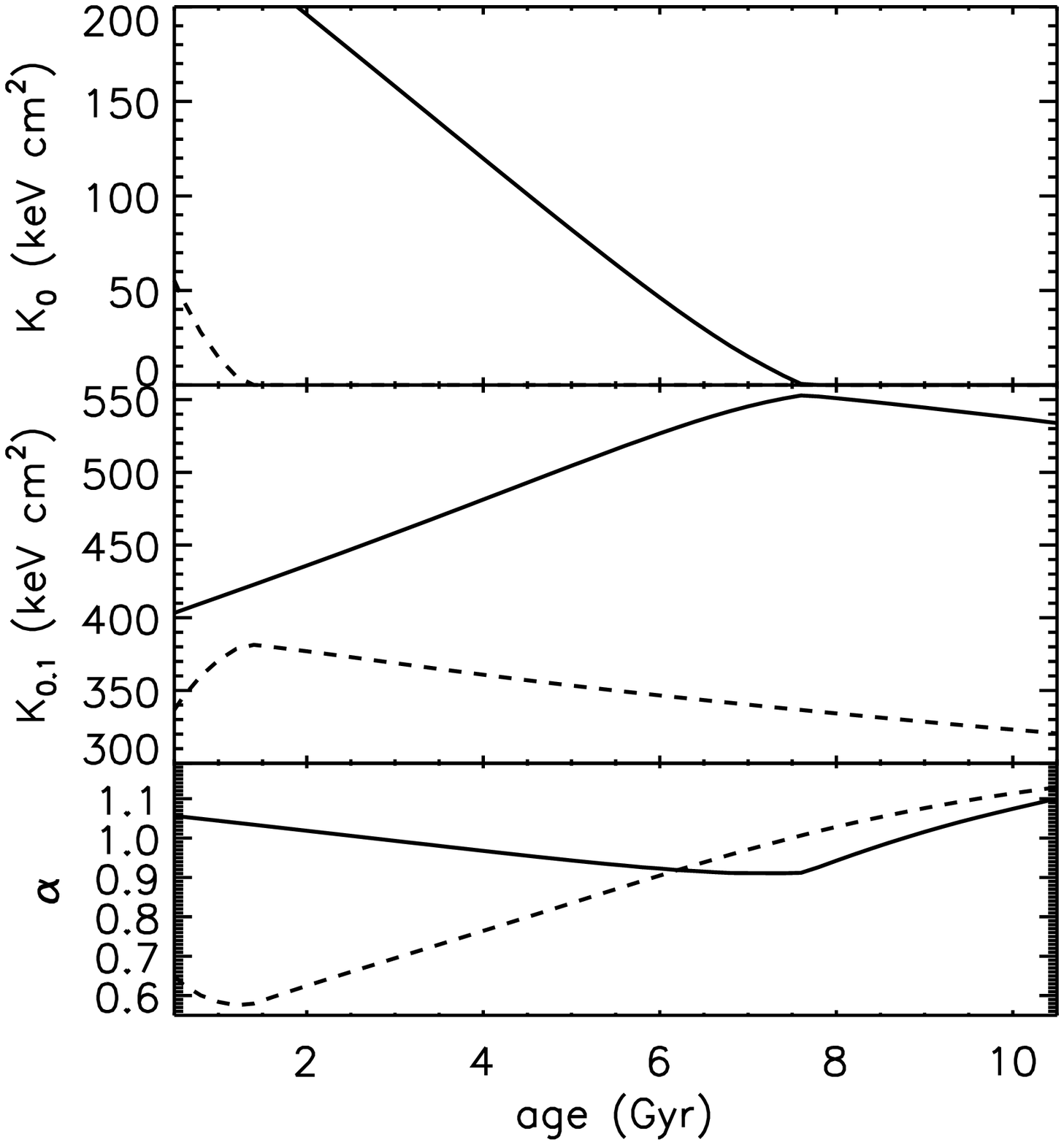,width=0.33\textwidth}
 \epsfig{figure=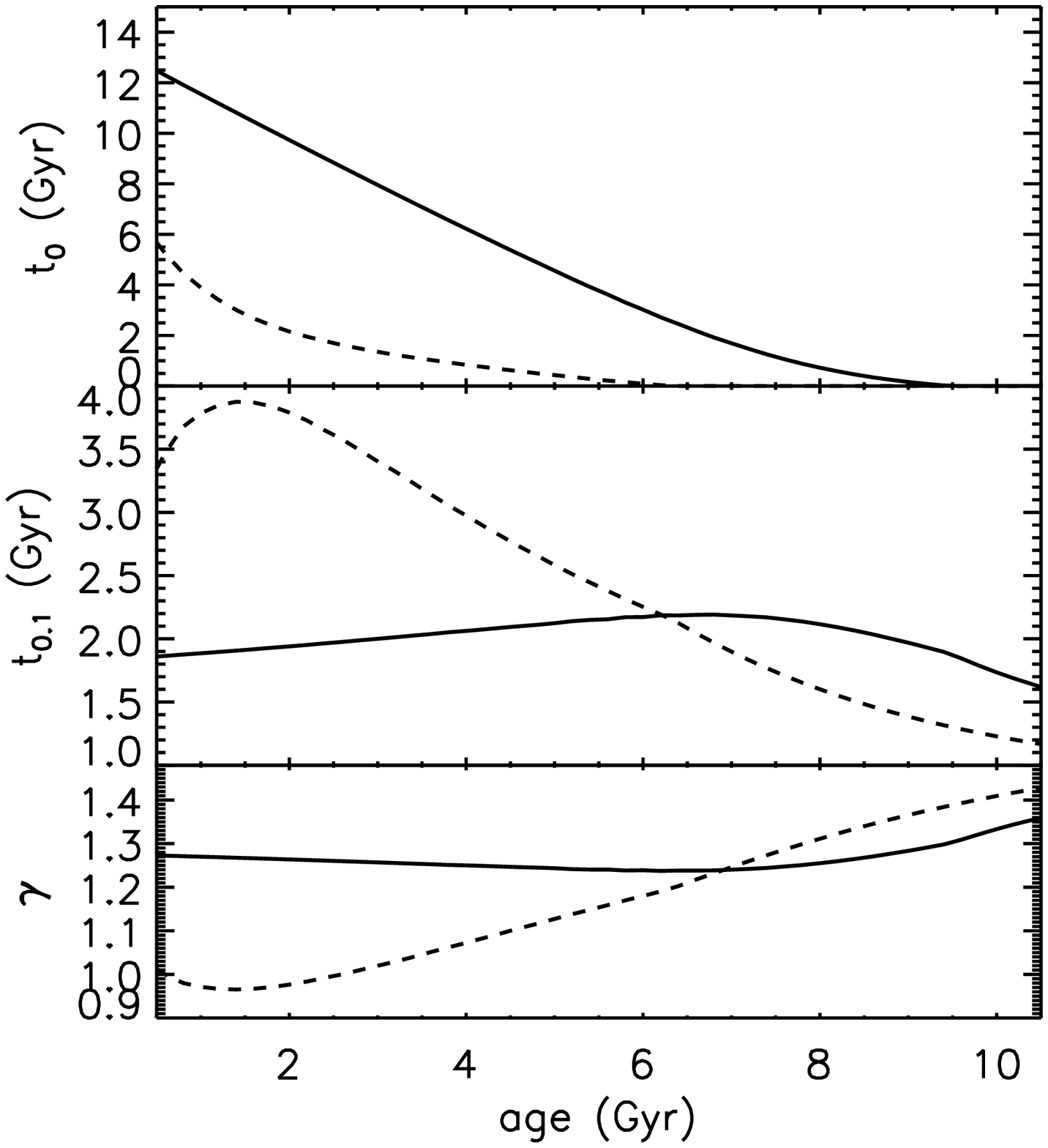,width=0.33\textwidth}
}
\caption{Properties of the cool cores as function of the clusters age
for the ``4 keV'' ({\it dashed line}) and the ``8 keV'' object ({\it solid line}).
({\bf Upper panels}) Evolution with the cosmic time of few
significant best-fit parameters in Eq.~\ref{eq:fit} (see text in Sect~3.2).
({\bf Lower panels}) Evolution of the cooling radius defined as the radius
at which the cooling time equals the Universe age at given redshift
({\it left}); distribution of the best-fit parameters
of the entropy profile ({\it middle}; cf. Eq.~\ref{eq:entr}) and
of the cooling time profile ({\it right}; see Eq.~\ref{eq:tcool}).
} \label{fig:tc_r}
\end{figure*}

\subsection{Modelization of the radial profiles}

To characterize the evolution of the cooling core,
we consider the results of the modelization of the radial profiles
of the X-ray quantities as function of the cosmic time.

We model the gas temperature, density and surface brightness profiles
shown in Figure~\ref{fig:prof} with analytic formulae commonly 
used in the literature, which provide a good description
of both observed and simulated profiles:
\begin{eqnarray}
T_{\rm gas}  & = & a_2 \frac{1 + x_0^{a_3}}{(1 + x_1^2)^{a_4}}  \nonumber \\
n_{\rm gas}  & = & b_2 x_0^{-b_3} \left(1 + x_1^2 \right)^{-1.5 b_4} \nonumber \\
S_{\rm b} & = & c_2 x_0^{-c_3} \left( 1 + x_1^2 \right)^{0.5-3 c_4} 
\label{eq:fit}
\end{eqnarray}
where $x_0$ and $x_1$ are the radius rescaled for the corresponding 
scale radius (e.g., $x_0=r/a_0$ or $r/b_0$ or $r/c_0$).
To avoid some degeneracy among the parameters describing the temperature 
profile, we fix the inner scale radius, $a_0$, to $0.05 R_{200}$.  
A least squares fit is then performed between $0.01 R_{200}$ and $0.5 R_{200}$, 
where the numerical simulations provide more robust results, propagating a
relative error comparable to the observational constraints (without
any significant change in the best-fit results, we have adopted errors
in the range of few per cent on the surface brightness, 
5-15 per cent on the gas density, 10-20 per cent on the temperature values).

In general, single power-laws provides a good description
of these profiles over partial sections of the radial profile, suggesting
that cooling alone is not able to reproduce an emission shaped with an inner,
well-defined core.
For instance, while a single $\beta-$model well reproduces the outskirts
of the gas density and surface brightness profile, 
a second inner component, either a power-law as in equation.~\ref{eq:fit} or
an additive $\beta-$model, is characterized by very small core radii,
of the order of the lower end of the investigated radial range of 
$0.01 R_{200}$, and a factor between 5 and 10 lower than the external
core radius.  

A power-law $+$ constant describes properly the 
gas entropy profile, $K=T_{\rm gas}/n_{\rm gas}^{2/3}$,
\begin{equation}
K = K_0 +K_{0.1} \left( \frac{r}{0.1 R_{200}} \right)^\alpha,
\label{eq:entr}
\end{equation}
and the cooling time profile, 
\begin{equation}
t_{\rm cool}=t_0 +t_{0.01} \left( \frac{r}{0.01 R_{200}} \right)^\gamma, 
\label{eq:tcool}
\end{equation}
over the radial range $0.01-0.1 R_{200}$, as we discuss in the following
subsection.

\subsection{Best-fit parameters and age of the structure}

We find that all the parameters that describe the structure of the cluster core
are strongly correlated with its age.
When a single power-law is fitted in the inner regions,
the slope in the gas temperature profile
starts from a value of zero (due to the initial isothermal assumption) and
reaches $\sim0.40$ at $z=0$ in the cool and hot system, respectively, 
under the action of the increasing radiative cooling.
Once equation~\ref{eq:fit} with $a_0=0.05 R_{200}$ is fitted, 
the steepening of the inner gradient is well represented by the parameter $a_3$, 
that rises from 0.1 up to 1.2 as the cooling core evolves (see Fig.~\ref{fig:tc_r}). 
At the same time, $a_1$ decreases significantly with the age from $0.4-0.5 R_{200}$
to values around $0.13 R_{200}$.

The gas density profile steepen in the center with the age 
of the structure and $b_3$ varies from 0.3 to 1.1
in the hot system, from 0.4 to 1.2 in the cool one (Fig.~\ref{fig:tc_r}).
On the other hand, the outer slope parameter, $b_4$, when a single
$\beta-$ model is fitted (i.e. $b_3$ is fixed equals to zero)
remains almost unchanged around values
of 0.48 and 0.54 in the less and most massive object, respectively.

Single power-laws fitted between
$0.01$ and $0.1 R_{200}$ and between $0.1$ and $0.5 R_{200}$,
show a clear steepening in the surface brightness profiles, 
with the inner slope that smoothly changes with time from $-0.6$ to $-1.2$
(down to $-1.4$ in the cool system) 
and the outer slope, 
that varies mildly from $-1.8$ to $-2$ in the more massive cluster,
and from $-1.7$ to $-1.8$ in the less massive system. 
When a single $\beta-$model is fitted
the outer slope parameter, $c_3$, is about 0.65 and 0.57 in the ``8 keV'' and
``4 keV'' object, respectively,
for $r>0.2 R_{200}$, whereas decreases from 0.5 to 0.4 (cluster)
and from 0.44 to 0.34 (group) when the joint-fit with the inner power-law
is performed. 

The logarithmic slope of the entropy profile $\alpha$ lies between
$0.9$ and $1.1$ in the hot cluster and increases slightly from
$0.6$ to $1.1$ in the cool system.
The pedestal value $K_0$ show the largest variation, from about 
270 (in the ``8 keV'' cluster; 100 in the ``4 keV'' one) keV cm$^2$ to zero in
$\sim 7.5 (2)$ Gyr (Fig.~\ref{fig:tc_r}).  

The slope $\gamma$ of the cooling time profile remains 
approximately constant around $1.3$ in the most massive system. It 
increases from $1.1$ to $1.4$ in the cool system with 
a corresponding decrease in the cooling time at $0.1 R_{200}$
from 4 to 1 Gyr (see Fig.~\ref{fig:tc_r}).
The pedestal value $t_0$ is again
the parameter more sensitive on the evolution of the central core, 
decreasing from 13.5 Gyr to zero in 9.5 Gyr in the massive object,
and from 8.5 Gyr to zero in 6 Gyr in the ``4 keV'' system.

Overall, the best-fit parameters that correlate more significantly 
with the age of the structure and the evolving cooling core 
are the inner slopes $a_3$, $b_3$, $c_3$ and the pedestal values
$K_0$ and $t_0$ (see Fig.~\ref{fig:tc_r}).
For instance, under the action of the radiative cooling alone, 
we predict a variation by a factor of 3 
in the inner slope of the gas density and surface brightness profile
of massive systems and a corresponding steepening of the cooling time
and entropy profiles within $0.1 R_{200}$.

The cooling radius itself, $R_{\rm cool}$, defined as the radius 
at which the cooling time equals the Universe age at given redshift, 
marks clearly the age of the structure subjected to radiative losses only,
evolving from $\sim 0.01 R_{200}$ at $z>2$ to $\sim 0.05$ and $0.06 R_{200}$
at $z=0$ in the 8 and 4 keV system (panel at the bottom right in 
Fig.~\ref{fig:tc_r}).

\begin{figure*} 
\hbox{
 \epsfig{figure=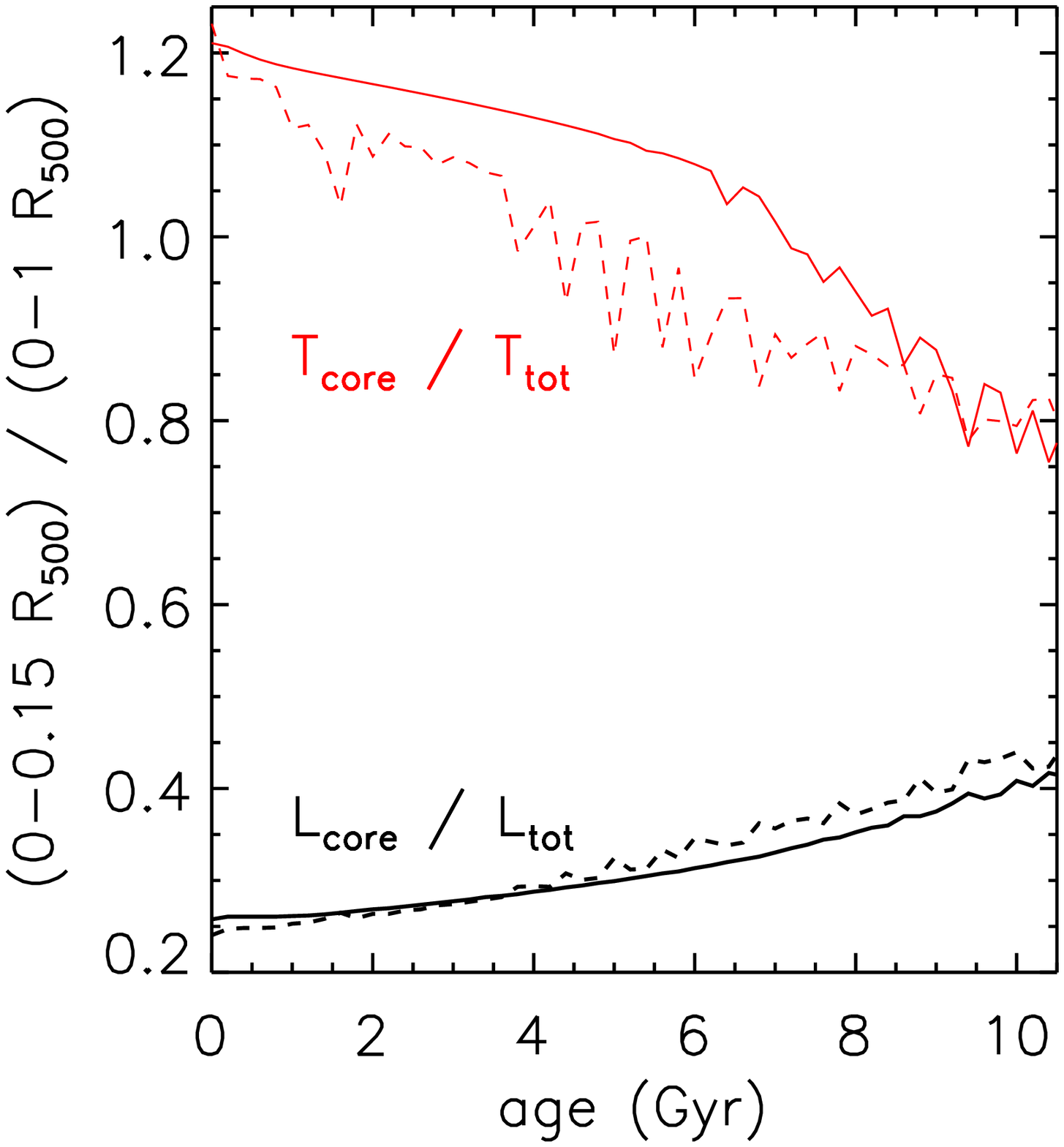,width=0.33\textwidth} 
 \epsfig{figure=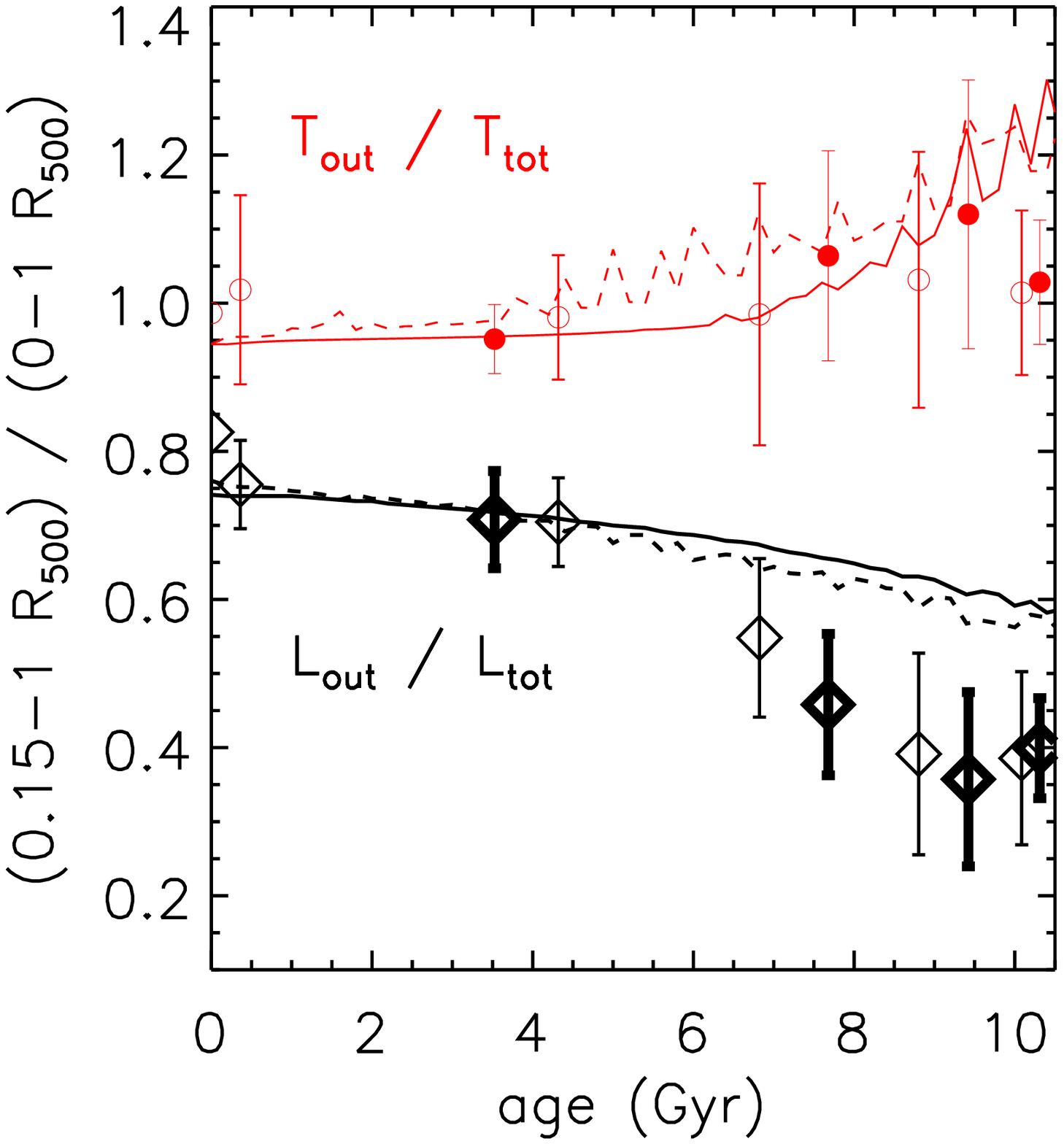,width=0.33\textwidth} 
 \epsfig{figure=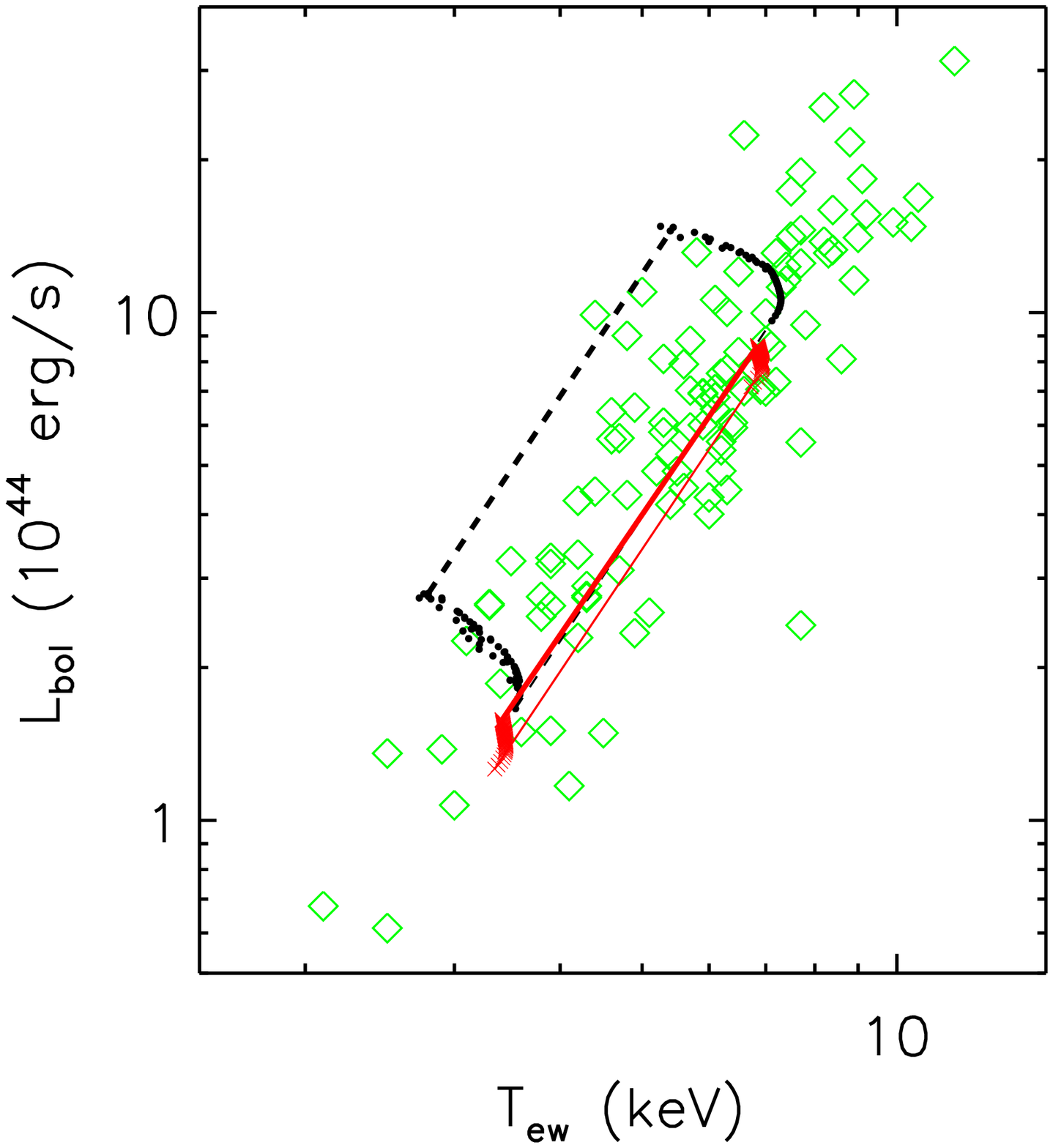,width=0.33\textwidth}
}
\hbox{
 \epsfig{figure=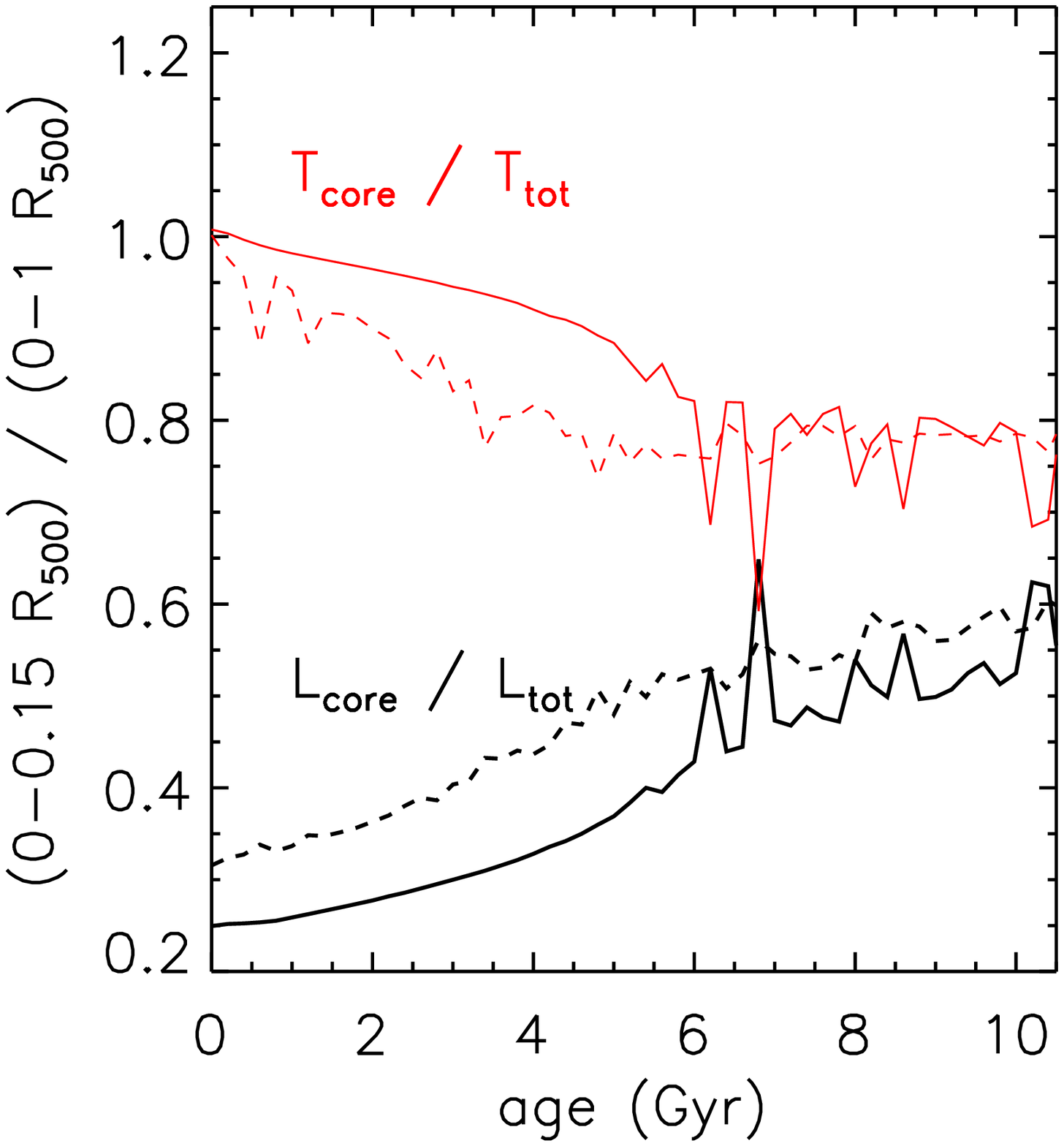,width=0.33\textwidth} 
 \epsfig{figure=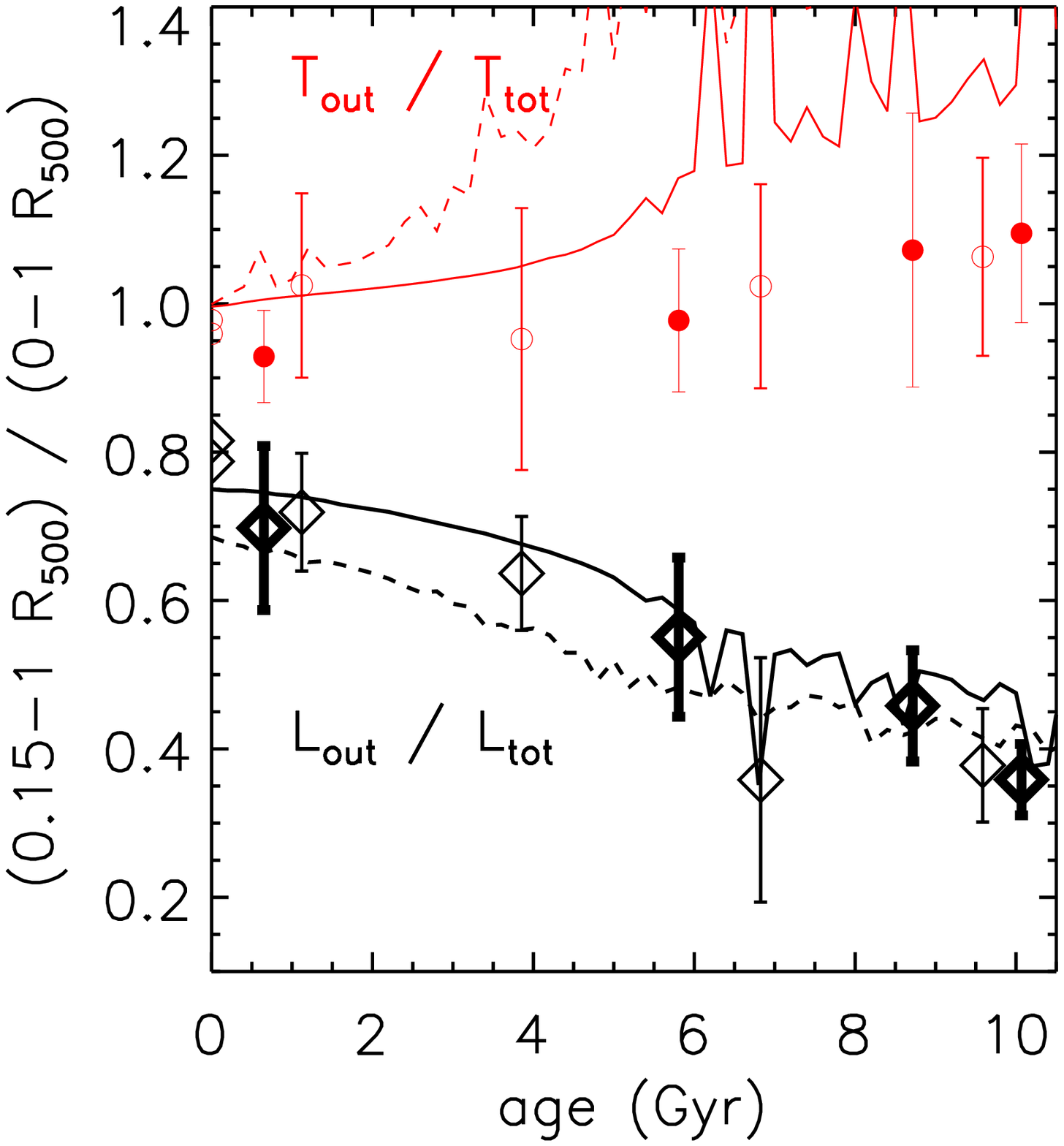,width=0.33\textwidth} 
 \epsfig{figure=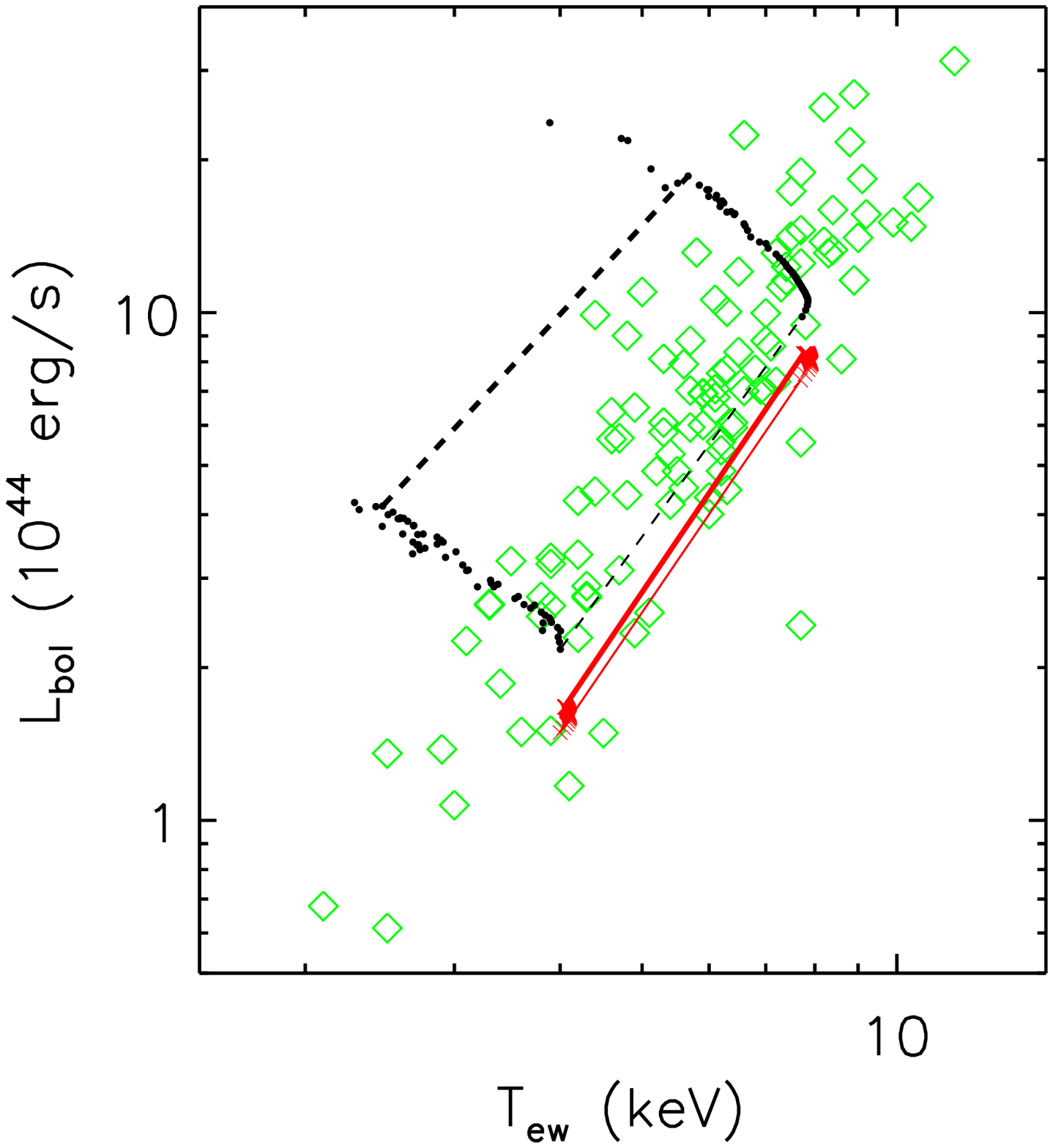,width=0.33\textwidth}
} 
\caption{Evolution of the gas temperature and X-ray bolometric luminosity measurements.
The {\bf upper panels} refer to the reference model, whereas the {\bf lower panels}
adopt an isothermal profile as initial condition as described in Sect.3.4.
({\bf Left}) Variation with the cosmic time of the core contribution to the total
measurements for the ``cool'' system ({\it dashed lines}) and the ``hot'' cluster
({\it solid lines}).
({\bf Middle}) Ratios between the values estimated at $0.15-1 R_{500}$ and within
$R_{500}$ compared with the observational constraints in Maughan et al. (2007).
The {\it thickest} diamonds represent the subsample of the most relaxed, less
elliptical objects, selected with respect to the median values in centroid shifts
and ellipticity measurements (see Table~2 in Maughan et al. 2007).
The age of the observed clusters is defined as the age that minimizes
the deviations between the observed and the predicted ratios in temperature
and luminosity.
({\bf Right}) Evolution in the $L-T$ diagram. We show estimates 
in the radial range $0.15-1 R_{500}$ ({\it red crosses}) and
within $R_{500}$ without the exclusion of any core ({\it dots}).
The lines connect the initial values ({\it thin line}) and the predicted ones
after 10 Gyrs of evolution ({\it thick line}).
The {\it green diamonds} represent the observed values in Maughan et al. 
(2007), when the core region $r<0.15 R_{500}$ is excised.
} \label{fig:lt} \end{figure*}

\subsection{Comparison with observational constraints}

The cool cores observed in nearby X-ray luminous galaxy clusters
(e.g. Donahue et al. 2006, Sanderson et al. 2006, Dunn \& Fabian 2008)
show radial profiles well described by a single power-law
with constant slope, like the inner temperature profile
($T_{\rm gas} \propto r^{0.4}$), the cooling time profile
($t_{\rm cool} \propto r^{1.3}$) and the entropy profile 
($K \propto r^{1.1}$) outside the cooling region. 
All these slopes are well matched in our simple models
(see, e.g., Fig.~\ref{fig:tc_r}).

We compare our X-ray luminosity and emission weighted
temperature estimates with the recent observational constraints obtained
for a sample of 111 galaxy clusters in the redshift range $0.1<z<1.3$ with
archived {\it Chandra} exposures in Maughan et al. (2007).
We estimate ``spectroscopic-like'' temperature $T_{sp}$ and luminosity 
both including and excluding
the core emission within $0.15 R_{500} \approx 0.1 R_{200}$.
The cool and the hot system behave similarly,
with a luminosity from the core that rises with time from 25 up to 40 
per cent of the total (within $R_{500}$) X-ray bolometric value, whereas the 
ratio between the emission weighted temperature in the core and within
$R_{500}$ decreases by about 35 per cent in 10 Gyrs.
When the temperature and the luminosity are calculated within $R_{500}$,
including the core region, the clusters evolve along wide trajectories
in the $L-T$ plane (see dots in the right panel of Fig.~\ref{fig:lt}).
The value of the slope $B$ of the $L-T$ relation remains constant 
around a value of 2.5, whereas the
normalization $A = (L/10^{44}$ erg s$^{-1}) / (T/5$ keV$)^B$
increases from 4.0 to 11.9 after 10 Gyrs.
On the other hand, if, instead of the total cluster emission, we
consider only the contribution from the region outside the core
(i.e. $0.15 R_{500} < r < R_{500}$), the luminosity and temperature
estimates do not change significantly with the cosmic time,
with a slope $B \sim 2.5$ and a normalization that rises
by 17 per cent only in 10 Gyrs from the initial value $A=3.4$.
The predicted $L-T$ relation is compared in the left panel of Fig.~\ref{fig:lt}
with the observational results obtained similarly from
Maughan et al. (2007). By using a $\chi^2$ minimization 
of a linear fit to the logarithmic values of the observed
luminosity and temperature estimated for 111 clusters in the radial range
$0.15 R_{500} - R_{500}$, we measure a slope of $2.77 \pm 0.08$, 
quite consistent with the range found in our simulated objects, and
a normalization $A = 4.26 \pm 0.11$, that is just a factor 1.07--1.25
larger than our simulated relation calculated excluding the inner
$0.15 R_{500}$ at the present time.

In Fig.~\ref{fig:lt}, the ratios between the gas temperature and luminosity
estimated in $0.15-1 R_{500}$ to those estimated in $0 - R_{500}$ are compared 
with the corresponding values estimated in the sample of Maughan et al. (2007).
In particular, we associate an age to each of the observed cluster, defined as
the age of the models that minimizes the squared differences between the observed 
and predicted ratios $T_{\rm out}/T_{\rm tot} = T_{0.15-1 R_{500}} / T_{0-1 R_{500}}$ 
and $L_{\rm out}/L_{\rm tot} = L_{0.15-1 R_{500}} / L_{0-1 R_{500}}$, weighted
by the propagated errors. 
Admittedly, this estimate for the age is very uncertain.
An upper limit on the age is fixed equal
to the age of the Universe at the object's redshift minus one Gyr, to allow
for the formation of the structure.
We remind here that what we mean for ``age'' of the structure
is something close, in the observational view of the cluster hierarchical formation,
to the time elapsed since the last major merging able to affect the physical 
properties of the core, or even to destroy it.
In other words, our ``age'' can be related to an ``observational age''
by the time during which a cluster evolves ``passively''.

The data are then ordered by the estimated age, binned by 15--20 elements 
and plotted in the central panel of Fig.~\ref{fig:lt}.
Our rough estimates indicate that about 30 per cent of the clusters
in the observed sample have an age less than 2 Gyr, 20 per cent
are between 2 and 6 Gyrs old, the remaining 50 per cent is more than
6 Gyrs old, with 13 low-redshift objects with an age around 10 Gyrs.
Moreover, the distribution of the age with respect to the median redshift ($z=0.324$)
indicates that the dynamically young structures (age $< 6$ Gyrs) are preferentially
located at higher redshift (33 out of 54 objects), 
whereas about 60 per cent of clusters at lower redshift 
have an age higher than 6 Gyrs (35 out of 56 clusters).
While the predicted ratio between the outer and total temperature ranges between
1 and 1.2 as actually observed, the contribution of the outer region to the total 
luminosity is expected from the models to be larger than 0.6 at 
any age, whereas the observed ratios reach values around 0.4.
This ratio requires that the contribution in luminosity
from the core with respect to the outer region should be twice
than actually measured in our models (after 10 Gyrs,
we estimate $L (<0.15 R_{500}) / L (0.15-1 R_{500}) \approx 0.7$,
while a value $\sim 1.5$ would be required to match the observations),
suggesting that more favourable conditions for cooling
have to be provided (like different initial conditions; see next subsection)
to generate objects with similar integrated properties.

\subsection{Dependence of the profiles upon the initial conditions}

The initial conditions described in section 2 are necessarily
somewhat arbitrary. For our reference models, described above, we decided
to use as initial temperature profiles at large radii the average profile
observed by Vikhlinin et al. (2006). In the core region, instead, we
considered isothermal gas. A different obvious choice would be to
assume isothermal gas in the whole cluster (as in Brighenti \& Mathews
2003 and Guo \& Oh 2007). With this set-up, however, the observed negative
temperature gradient for $r > 0.2\,R_{500}$ is not reproduced. 

It is useful to compare how the results change if the ICM is assumed to be
initially isothermal at every radius, with temperature calculated
from the observed $M_{\rm tot}-T$ relation by Arnaud et al. (2005).
In this case, the central gas density required to have $f_{\rm b}\sim 0.16$
at the virial radius is larger than in our reference models. This implies
a correspondingly shorter cooling time (by $\sim 60$\%). 
As a consequence of that,
our initially isothermal clusters age faster
(the cooling catastrophe occurs earlier and reach a quasi steady-state
sooner).
At any given time, the cool cores are more developed than in the
reference models, with a steeper density and surface brightness
profiles with a ratio $n_{\rm gas} (0.01 R_{200})/n_{\rm gas} (0.2 R_{200})$ 
that reaches a value of 55 after 10 Gyr and the corresponding ratio 
in $S_{\rm b}$ more that a factor 2 larger than the reference value of 
$\sim 60$ (see Fig.~\ref{fig:dq_age}).
 
The relative contribution from the core to the total luminosity rises
with the age of the core up to a value of 0.6, fifty per cent higher than
maximum contribution observed in the reference models.
Even the $L-T$ relation changes significantly as consequence of the larger
emission from the core: when only the outer regions are considered (i.e.
$0.15-1 R_{500}$), the normalization $A$ varies around 2.7, 60\% lower
than the observed value in the Maughan et al. (2007) sample (see
Fig.~\ref{fig:lt}).

We conclude that our results depend significantly upon the initial 
temperature profile chosen, if we maintain that
gas fraction measured at the virial radius has to be consistent with
the cosmological value.
Once we adopt an initial temperature profile that well matches
the observational constraints at $r>0.15 R_{500}$, we generate 
systems with a cool core in formation and not fully developed. 
The reason for this inconsistency is likely to be the non-evolution 
of the potential well in our models. Real massive clusters are still
slowly growing with
time (e.g. Li et al. 2007) and in their youth the ICM would have been
cooler, with a shorter cooling time. This would contribute to develeop
a mature cool core sooner. To account the effects of an evolving
potential well is beyond the goal of the present paper.

\section{Discussion and Conclusions}

We present a simple gasdynamic modelization of the X-ray emitting plasma 
in galaxy clusters subjected to the action of radiative processes only. 
These toy--models allow to follow and quantify the variations as function
of the cosmic time of the physical quantities describing the cool cores
in systems with a virial mass of $1.3 \times 10^{14} M_{\odot}$ 
and $1.3 \times 10^{15} M_{\odot}$, typical of a ``4 keV'' cluster 
and a ``8 keV'' cluster.

The cooling radius, $R_{\rm cool}$, defined as the radius at which the cooling 
time equals the Universe age at given redshift, evolves from $\sim 0.01 R_{200}$ 
at $z>2$, where the structures begin their evolution, to $\sim 0.05 (0.06) R_{200}$ 
at $z=0$ in the 8 keV (4 keV) cluster (Fig.~\ref{fig:tc_r}).

The temperature decreases with a mean rate around 10 per cent per Gyr 
when measured at $0.01 R_{200}$, with the largest variation rate happening
in the hot cluster when it is $\sim 7$ Gyr old with a decrement
of the order of 15 per cent per Gyr.
At the same time, the gas density and surface brightness
increase with a rate of 15-20 per cent per Gyr at $0.01 R_{200}$, reaching an 
incremental rate of 8 per cent per Gyr at the present time.
As shown in the lower panels of Fig.~\ref{fig:prof}, the systems reach
a quasi-steady state. 
On the other hand, the cool system has the largest variation
at the beginning of its evolution (20 per cent in
$T_{\rm gas}$, 30 per cent in $n_{\rm gas}$ and $S_{\rm b}$) and, then,
varies slowly up to changes of few per cent after 10 Gyrs (Fig.~\ref{fig:dq_age}).
For pedagogical reasons, we run our models to a simulated
age of 30 Gyr. Considering the state of still-developing cool cores, 
the variations in the future are quite significant, with a predicted rise 
of 78 (54) and 161 (126) per cent in $n_{\rm gas}$ and $S_b$, respectively, 
in the hot (cool) system and a decrease of 18 (10) per cent in $T_{\rm gas}$
at $0.01 R_{200}$.

After 10 Gyr of radiative losses, the ratio between the gas temperature 
at $0.01 R_{200}$, well within the cooling radius, and at $0.2 R_{200}$, 
located beyond the cluster core, is equal to 0.4.
The ratio between the gas densities at the same radii is $\sim 35$,
a factor $\sim 4-5$ larger than the ratio measured at the beginning
of the formation of the cooling core.
The surface brightness is observed to increase with the age of the structure
with a mean rate of 20 and 2 per cent per Gyr at $0.01$ and $0.2 R_{200}$, respectively,
implying a ratio that rises from 10 to $\sim$60 at the present time in both of the 
objects.

The cooling time and gas entropy radial profiles are well represented by
power-law functions, $t_{\rm cool}=t_0 +t_{0.01} (r / 0.01 R_{200})^{\gamma}$
and $K=K_0 +K_{0.1} (r / 0.1 R_{200})^{\alpha}$,
with $t_0$ and $K_0$ that decrease with time from 13.5 Gyrs and 270 keV cm$^2$ in the hot
system (8.5 Gyr and 100 keV cm$^2$ in the cool one) and reaches zero after 9.5 (6) Gyrs.
The slopes vary slightly with the age, with $\gamma \approx 1.3$ and
$\alpha \approx 1.1$ (Fig.~\ref{fig:tc_r}).

The behaviour of the inner slopes of the gas temperature and density profiles
are the most sensitive and unambiguous tracers of an evolving cooling core.
Their values after 10 Gyrs of radiative losses, $T_{\rm gas} \propto r^{0.4}$ and
$n_{\rm gas} \propto r^{-1.1 (-1.2)}$ for the hot (cool) object,
are remarkably in agreement with the observational constraints available for nearby
X-ray luminous cooling core clusters (e.g. Donahue et al. 2006,
Sanderson et al. 2006, Dunn \& Fabian 2008; see also the case of A1835 in 
Fig.~\ref{fig:prof}). This implies that the heating process, if presently active,
must not alter the ``pure cooling flow'' profiles in cool core clusters.

The emission-weighted temperature diminishes by about 25 per cent and the 
bolometric X-ray luminosity rises by $\sim60$ per cent in 10 Gyrs when all the
cluster emission is considered in the computation. On the contrary, 
when the core region within $0.15 R_{500}$ is excluded, the gas temperature value
does not vary and the X-ray luminosity changes by $20$ per cent only.
Observational constraints on the distribution of the $T (0.15-1 R_{500}) / T (<R_{500})$
as function of redshift (Maughan et al. 2007) show a mean value around 1
suggesting that the observed clusters cannot be significantly older than
8 Gyrs, once compared with the predictions from our models, independently from
the initial conditions on the adopted temperature profile (see Fig.~\ref{fig:lt}).
As consequence of that, the value of the $L_{\rm out} / L_{\rm tot}$
ratio obtained in our model should be about 50 per cent lower
to match the lower end of the observed distribution, implying
a ratio between the core and the outer luminosities of $\sim1.5$, 
instead of the present predictions of about $0.4-0.7$ 
(see panel on the right in Fig.~\ref{fig:lt}).
More cooling at the core is therefore required to meet the observed trends.
We show that this is can be partially obtained by more flat temperature 
distribution in the outskirts, once the total gas fraction at the virial
radius is fixed. Therefore, a tension between predicted and 
observed properties rises however when both the temperature profile and 
the temperature--luminosity relation have to be recovered 
consistently.

In this prospective, more detailed models following the energy distribution 
in the cluster outskirts in a cosmological context are required 
and matter for a following study.

\section*{Acknowledgments}

We acknowledge the financial contribution from contract ASI-INAF I/023/05/0 and I/088/06/0.
We thank the anonymous referee for very useful comments that improved
the presentation of the work.

\end{document}